\documentclass{article}

\usepackage{arxiv}

\usepackage[utf8]{inputenc} 
\usepackage[T1]{fontenc}    
\usepackage{graphicx}
\usepackage{amssymb}
\usepackage{bbm}
\usepackage{amsmath}
\usepackage{hyperref}

\usepackage{xcolor}
\usepackage{algorithm2e}
\RestyleAlgo{ruled}

\newcommand\red[1]{{\color{black}{{#1}}}}
\newcommand{\bs}[1]{\boldsymbol{#1}}
\newcommand{\R}{\mathbb{R}}
\providecommand{\abs}{\mathrm{abs}} 

\title{Nonlinear blind source separation exploiting spatial nonstationarity}

\author{Mika Sipilä \\
	Department of Mathematics and Statistics\\
	University of Jyvaskyla\\
	Finland \\
	\And
	Klaus Nordhausen \\
	Department of Mathematics and Statistics\\
	University of Jyvaskyla\\
	Finland \\
        \And
        Sara Taskinen \\
	Department of Mathematics and Statistics\\
	University of Jyvaskyla\\
	Finland \\
}

\begin{document}

\maketitle

\begin{abstract}
In spatial blind source separation the observed multivariate random fields are assumed to be mixtures of latent spatially dependent random fields. The objective is to recover latent random fields by estimating the unmixing transformation. Currently, the algorithms for spatial blind source separation can only estimate linear unmixing transformations. Nonlinear blind source separation methods for spatial data are scarce. In this paper we extend an identifiable variational autoencoder that can estimate nonlinear unmixing transformations to spatially dependent data and demonstrate its performance for both stationary and nonstationary spatial data using simulations. In addition, we introduce scaled mean absolute Shapley additive explanations for interpreting the latent components through nonlinear mixing transformation. The spatial identifiable variational autoencoder is applied to a geochemical dataset to find the latent random fields, which are then interpreted by using the scaled mean absolute Shapley additive explanations. Finally, we illustrate how the proposed method can be used as a pre-processing method when making multivariate predictions.
\end{abstract}


\section{Introduction}
Nowadays spatially indexed data are encountered in many fields of science. For example, in geochemistry, samples are collected at different locations and analysed for their chemical compositions to detect patterns, for identifying areas of pollution, or for mining purposes. From now on, we denote the location as $\bs s \in \mathcal S \subset \R^q$, and let $\bs x(\bs{s}) =(x_1(\bs{s}),\ldots,x_d(\bs{s}))^\top$ be the observable $d$-variate vector. Here the set of possible locations $\mathcal S$ is the domain and the data are usually called spatial data. In most applications $q=2$ and for simplicity, without loss of generality, we also assume the same from now on. Spatial data are usually characterized by its mean function $\bs m(\bs s)$ and a (spatial) covariance function $\bs C(\bs x(\bs s),\bs x(\bs s'))= E\left((\bs x(\bs s)-\bs m(\bs s))(\bs x(\bs s')-\bs m(\bs s'))^\top\right)$, where $\bs s, \bs s' \in \mathcal S$. The main idea in modeling  is that observations closer together are more similar than observations further apart. However, as the covariance function is usually quite complex, modeling spatial data is often quite difficult and is, for example, much more challenging than modeling time series data, which have a natural direction (past to present) that is missing in spatial data. Thus, $\bs x(\bs s)$ is quite often assumed as stationary to simplify spatial modeling, meaning that $\bs m(\bs s) = \bs m$ for all $\bs s \in \mathcal S$ and  $\bs C(\bs x(\bs s),\bs x(\bs s'))= \bs C(||\bs s-\bs s'||)$ for all $\bs s,\bs s' \in \mathcal S$, i.e. the mean is the same at all locations and the spatial dependence depends only on the distance between observations. For an overview  discussing the complexity of modeling $\bs C$, see \cite{GentonKleiber2015}.

Another suggestion for simplifying spatial data modeling is considering a latent component modeling as a pre-processing step. The data are then represented using $d$ independent, latent components in $\bs z(\bs s) = (z_1(\bs{s}),\ldots,z_d(\bs{s}))^\top$. Thus, after pre-processing, univariate models can be fitted to components and the components can be interpreted individually and predicted separately \cite{MuehlmannNordhausenYi2021}. \red{While the latent components can simply be seen as useful modeling tools, they often turn out to have physical meanings for which however subject knowledge is usually needed}. A popular approach to latent component modeling is linear blind source separation (BSS) \cite{comon2010handbook}, which aims to find independent latents components using the observed data by assuming that the observations are generated from the components through some linear mixing system. Recently, \cite{NordhausenOjaFilzmoserReimann2015,bachoc2018spatial,MuehlmannBachocNordhausenYi2024} suggested a spatial BSS (SBSS) approach for stationary multivariate spatial data,  where the independent latent components are obtained by jointly diagonalizing two or more moment-based matrices. SBSS was also extended to nonstationary multivariate spatial data by \cite{MuehlmannBachocNordhausen2021}, then yielding nonstationary spatial source separation (SNSS). In the SNSS model, $\bs x(\bs s)$ is assumed to be stationary with respect to $\bs m$ but not with respect to $\bs C$. A drawback of SBSS and SNSS is that both assume a linear mixing procedure, which is interpretable and mathematically tractable but might be too  simplistic in reality.

The motivation for the use of nonlinear BSS approach is the same as for linear BSS, that is, one wants to find a useful representation of high-dimensional data to be used in further analyses \cite{Hyvarinenetal:2023}. As the mixing procedure is assumed to be nonlinear, the components are unidentifiable https://www.overleaf.com/project/637f3b06f58fac80b0f38431without additional assumptions on the mixing function or on the distribution of the latent components \cite{hyvarinen1999nonlinear}. Recently, unsupervised deep learning methods, such as variational autoencoders (VAEs) \cite{kingma2013auto}, and their extensions, such as \cite{higgins2017betavae, zhao2017infovae}, were applied to find unmixing function and latent components based on the observed data. 
As the regular VAE solutions in general suffer from unidentifiability issues,
\cite{Khemakhem2020} proposed in a time series context the identifiable VAE (iVAE) method, which uses some additionally observed data to make the VAE model identifiable. In \cite{halva2021disentangling} a structured nonlinear BSS framework, which is suitable also for two dimensional graphs, was proposed. However, to our knowledge, models for spatial nonlinear BSS have not been developed or studied yet.

In this paper, we extend the iVAE method for spatially dependent data and study its performance using vast simulation studies and a real geochemical data example. Nonlinear BSS framework in general lacks tools for latent component interpretation, which is an essential task when analyzing the discovered latent representation. Therefore, a procedure based on Shapley additive explanations \cite{kernelshap} for interpreting the latent components through nonlinear mixing transformation is introduced and illustrated. The paper is organized as follows. We review the basics of linear and nonlinear BSS in Section \ref{sec:background}. Sections \ref{sec:SNSS-iVAE} and \ref{sec:scaled_mashap} introduce new iVAE method and new tools for interpreting the latent components, respectively. In Sections~\ref{sec:simulations} and \ref{sec:example}, simulation studies and a real data example are used to illustrate the performance properties of iVAE. The paper is concluded with a discussion in Sections~\ref{sec:conclusions} and \ref{sec:discussion}.

\section{Background}
\label{sec:background}

Before showing how iVAE can be extended to spatial data, we give some background on linear and nonlinear BSS as well as Shapley values, which are later used to interpret the latent components. 

\subsection{Linear and nonlinear BSS}
\label{sec:background_1}

Assume for now that $\bs x=(x_1, \dots, x_d)^\top$ is a $d$-variate observable random vector. In linear blind source separation (BSS) one assumes that $\bs x$ is a linear mixture of unknown $d$-variate latent components $\bs z=(z_1, \dots, z_d)^\top$, that is, 
\begin{align} \label{ICAmodel}
    \bs x=\bs A\bs z,
\end{align}
where $\bs A$ is an invertible $d\times d$ mixing matrix. The goal in BSS is to recover both $\bs A$ and $\bs z$ based on $\bs x$ only, but this cannot be done without making some additional assumptions on $\bs z$. Different BSS methods vary in the assumption made on latent components. For example, in independent component analysis (ICA) we assume iid data with independent latent components, and in second-order source separation (SOS) we assume that the components are weakly stationary time series. For general overview of BSS methods, see \cite{comon2010handbook}. In \cite{bachoc2018spatial}, spatial BSS methods were introduced. 

As the assumption on linear mixing is often too restrictive, nonlinear BSS methods have been introduced. For a recent review of methods, see \cite{Hyvarinenetal:2023}. Notice that although the methodology is often referred to as nonlinear ICA, in what follows, we use the more general term, nonlinear BSS, as it allows the components to be of any data type.
In nonlinear BSS one assumes that $\bs x$ is generated by applying an unknown invertible mixing transformation $\bs f:\mathbbm{R}^d\to\mathbbm{R}^d$ on the latent components $\bs z$ as 
\begin{align}
    \bs x = \bs f(\bs z).
\label{eq:nonlinearICA}
\end{align}
The objective is then to identify the transformation $\bs q:\mathbbm{R}^d\to\mathbbm{R}^d$ which returns $\bs z$ as
\begin{align}
    \bs z = \bs q(\bs x)
\label{eq:unmixing}
\end{align}
based on the observations of the vector $\bs x$ only. Generally, \red{without additional assumptions on the mixing function or on the distribution of the latent components}, nonlinear BSS is unidentifiable as there is an infinite amount of nonlinear transformations to generate mutually independent components from the observations \cite{hyvarinen1999nonlinear}. Thus, additional constraints on the distribution of $\bs z$ must be introduced to obtain identifiability without restricting the mixing transformation $\bs f$ severely.

 In many recent studies \cite{HyvarinenMorioka2016, HyvarinenMorioka2017, Hyvarinen2019, Khemakhem2020, HalvaHyvarinen2020}, the main assumption leading to identifiability is that each latent component $z_i$, $i=1,\dots,d$, is statistically dependent on an $m$-dimensional auxiliary variable $\bs u$, and that the conditional distribution is a member of the family of exponential distributions with parameters $\lambda_{i,k}$, $k=1,\ldots, r$. 
Assuming the independence of the $d$ conditional latent distributions yields then the joint distribution $p(\bs z|\bs u) = \prod_{i=1}^d p(z_i|\bs u)$. Following  \cite{Khemakhem2020}, the dependence of the parameters $\lambda_{i,k}$ on $\bs u$ is expressed using a function $\bs \lambda(\bs u)=(\bs \lambda_1(\bs u),\ldots,\bs \lambda_d(\bs u) )$, where $\bs \lambda_i(\bs u)=(\lambda_{i,1}(\bs u), \dots, \lambda_{i,r}(\bs u))^\top$ contains the parameters depending on $\bs u$ for component $i$. This function must be learned, for example, using some neural network. The joint conditional distribution can then be written as 
\begin{align}
    p_{\bs T, \bs \lambda}(\bs z| \bs u) = \prod_{i=1}^d \frac{Q_i(z_i)}{Z_i(\bs u)} \text{exp}\left[ \sum_{k=1}^r T_{i,j}(z_i) \lambda_{i,k}(\bs u) \right],
    \label{eq:sourcedist}
\end{align}
where $Q_i(z_i)$ is a base measure, $Z_i(\bs u)$ is a normalizing constant, and $\bs T_i(z_i)=(T_{i,1}(z_i) , \dots, T_{i,r}(z_i))^\top$ contains sufficient statistics. The dimension $r$ of each sufficient statistic $\bs T_i(z_i)$ and $\bs \lambda_i(\bs u)$ is assumed to be fixed. By assuming the source distribution (\ref{eq:sourcedist}) and the additionally observed auxiliary variable $\bs u$, the aim becomes identifying the latent sources using the unmixing transformation $\bs q(\bs x, \bs u)$. 

In a time series context there are several examples of auxiliary variables found in the literature. In the case of stationary time series, one can use as $\bs u$, for example, the previous observations \cite{HyvarinenMorioka2017}, and for nonstationary time series data, the time segment of the observation can be used as \cite{HyvarinenMorioka2016,Khemakhem2020,HalvaHyvarinen2020}.
We propose in Section \ref{sec:SNSS-iVAE} a nonlinear BSS method for spatial data. The method extends the method proposed in \cite{Khemakhem2020} that uses VAEs with auxiliary variables to learn the full identifiable generative model, and is thus referred as identifiable VAE (iVAE). 

\subsection{SHAP values and mean absolute SHAP values}
\label{sec:background_2}

The main advantage of using nonlinear BSS methods is that they can discover some meaningful underlying latent representation through a nonlinear mixing environment \red{by recovering the true latent components $\bs z$}. However, the interpretation of the latent representation can be difficult without knowing how the observed variables interact with the latent components. In case of linear BSS, it is easy to inspect the contributions through mixing/unmixing matrices, but when the mixing function is nonlinear, we do not have such methods available. Later in this paper, we suggest \red{novel metrics for evaluating contributions of input variables for nonlinear models with multiple outputs, such as iVAE. We explain the procedure of calculating these values in detail in Section~\ref{sec:scaled_mashap}, but let us first review here the Shapley values \cite{shapley1953value} and Shapley additive explanations (SHAP) \cite{kernelshap} that our metrics base upon.}

The SHAP framework is based on the Shapley values, which are originally used in cooperative game theory to fairly distribute the payout of a game between a group of players. Let $\mathcal{K}$ be a set of $K$ players and $\gamma$ be a function giving the payout of the game for any set of players $K$. The Shapley values calculated based on the function $\gamma$ produce the attributions $a_i$ for each player $x_i \in \mathcal{K}$ so that 
\begin{align}
    \gamma(A) = a_0 + \sum_{i = 1}^K a_i x_i',
\end{align}
where $A \subseteq \mathcal{K}$, $a_0$ is a baseline output without any players and $x_i' = \mathbbm{1}(x_i \in A)$ is a binary variable which gets a value of 1 if $x_i$ belongs in set $A$ and value of 0 otherwise. The function $\mathbbm{1}$ is an indicator function.
The Shapley value for player $x_i$ is then given by
\begin{align}
    a_i = \sum_{A \subseteq \mathcal{K} \setminus x_i} \frac{ |A|! (|\mathcal{K}| - |A| - 1)!}{K!}(\gamma(A \cup x_i) - \gamma(A)),
\label{eq:shapley_value}
\end{align}
where $|\cdot|$ denotes the size of a set.
The game can be looked as any mathematical function. Then, the function inputs correspond the players, and the output of the function corresponds the payload.
If the game $\gamma$ is a statistical model, the players are observations of $K$ variables $\mathcal{K} = \{ x_1, \dots, x_K \}$ and the payout is the predicted value of a variable $y$ produced by $\gamma$ with an input $A \subseteq \mathcal{K}$, then the attribution of an observation $x_i$ of $i$th variable to the outcome prediction is given by $a_i$.

Typically, statistical models have fixed input size, which is why the formulation in (\ref{eq:shapley_value}) is not directly applicable. The SHAP framework suggests a procedure where $\gamma(A)$ is estimated with $E(\tilde{\gamma}(\bs x) | A)$, where $\bs x = (x_1, \dots, x_K)^\top$ and $\tilde{\gamma}$ has a fixed input size of $K$. The Shapley values calculated based on these conditional expected values $E(\tilde{\gamma}(\bs x) | A)$ are called SHAP values, mathematically given by
\begin{align}
    \tilde{a}_i = \sum_{A \subseteq \mathcal{K} \setminus x_i} \frac{ |A|! (|\mathcal{K}| - |A| - 1)!}{K!}(E(\tilde{\gamma}(\bs x) | A \cup x_i) - E(\tilde{\gamma}(\bs x) | A)).
\label{eq:shap}
\end{align}
The conditional expected value $E(\tilde{\gamma}(\bs x) | A)$ is calculated in practice by averaging over the outputs of $\tilde{\gamma}$ of a representative background data while keeping the observed values in set $A$ fixed. While it is possible to calculate the exact SHAP values using this procedure, the computational burden becomes heavy when the number of variables is high. To reduce the computational burden, the kernel SHAP method \cite{kernelshap} can be used. The kernel SHAP approximates the SHAP values $\tilde{a}_i$ based on local interpretable model-agnostic explanations (LIME) \cite{LIME2016} by minimizing the equation
\begin{align}
    \zeta &= \sum_{A \subseteq \mathcal{K}} \left( E(\tilde{\gamma}(\bs x) | A) - \tilde{a}_0 + \sum_{i = 1}^K \tilde{a}_i x_i'  \right)^2 \pi(A),
\end{align}
where
\begin{align}
    \pi(A) = \frac{K - 1}{{K \choose |A|} |A| (K - |A|)},
\end{align}
with respect to the values $\tilde{a}_i$.  By using the specific kernel $\pi(A)$ called Shapley kernel, the algorithm in fact recovers the Shapley values. For more details, see \cite{kernelshap}.

The SHAP values are calculated individually for each (multivariate) observation/prediction pair making them a useful tool to interpret how each component of the observation affects the model prediction. In many cases we are interested in the population level effects, that is, how the observed variables affect the model prediction on population level. Methods, such as mean absolute SHAP (MASHAP) \cite{marcilio2020explanations} and Shapley global additive importance (SAGE) \cite{sage}, aim to obtain the population level importance for the input variables. In this paper we focus on MASHAP values, \red{which can be easily extended to models with multiple outputs as will be shown in Section~\ref{sec:scaled_mashap}}. Let now $n$ be the number of observations and $d$ the number of observable variables, and let $\tilde{a}_{i,j}$ be a SHAP value for $i$th variable of $j$th observation. The MASHAP value for the $i$th variable is then calculated as 
\begin{align}
    v_i = \frac{1}{n} \sum_{j = 1}^n |\tilde{a}_{i,j}|.
\label{eq:mashap}
\end{align}
\red{The MASHAP values $v_i$ can be interpreted as contribution or importance of $i$th input variable to the whole output population.} 

\section{Nonstationary spatial source separation using VAE}
\label{sec:SNSS-iVAE}

\begin{figure}[htb]
\centerline{\includegraphics[width=0.6\textwidth]{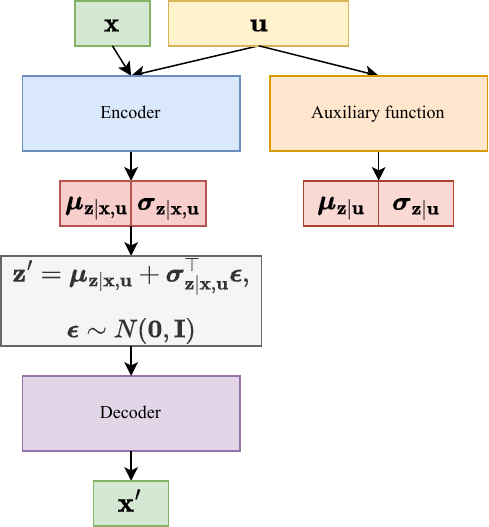}}
\caption{Schematic presentation of iVAE. iVAE learns simultaneously the encoder, the decoder, and the auxiliary function. \red{The latent components $\bs z$ are obtained as $\bs \mu_{\bs z|\bs x, \bs u}$, which is a mean of the variational distribution given by the encoder.}}
\label{fig:ivae}
\end{figure}

In this paper, we use iVAE \cite{Khemakhem2020} to solve the nonlinear SNSS problem \red{by assuming nonstationarity in the variance of the latent fields}. We use spatial segmentation as an auxiliary variable $\bs u$ and assume that the sources are distributed as in~(\ref{eq:sourcedist}). With spatial segmentation we mean that the domain $\mathcal S$ is divided into $m$ segments $\mathcal S_i \subset \mathcal S$, so that $S_i \cap S_j = \emptyset$ for all $i \neq j$ and $\cup_{i=1}^m \mathcal S_i = \mathcal S$, $i,j = 1,\dots,m$. Then, the auxiliary variable for the observation $\bs x(\bs s)$ is $\bs u(\bs s)=(\mathbbm{1}(\bs s \in \mathcal S_1), \dots, \mathbbm{1}(\bs s \in \mathcal S_m)))^\top$, i.e. a $m$-dimensional \red{standard} basis vector giving the segment corresponding to the spatial location of the observation. From now on, we use the notation $\bs x := \bs x(\bs s), \bs z := \bs z(\bs s)$, and $\bs u := \bs u(\bs s)$ for the sake of convenience. iVAE assumes the following generating model:
\begin{align}
    p_{\bs f,\bs T, \bs \lambda}(\bs x, \bs z|\bs u) = p_{\bs f}(\bs x|\bs z)p_{\bs T, \bs \lambda}(\bs z|\bs u),
\label{eq:generativemodel}
\end{align}
where $p_{\bs f}(\bs x|\bs z)$ is defined as
\begin{align}
    p_{\bs f}(\bs x|\bs z) = p_{\bs \epsilon_{\bs f}}(\bs x - \bs f(\bs z)),
\label{eq:f_eps}
\end{align}
which means that $\bs x$ can be composed of an independent noise vector $\bs \epsilon_{\bs f}$ and a mixing transformation $\bs f(\bs z)$ as $\bs x = \bs f(\bs z) + \bs \epsilon_{\bs f}$. The distribution $p_{\bs \epsilon_{\bs f}}$ is assumed to be a zero mean Gaussian distribution with infinitesimal variance to match the form of the nonnoisy nonlinear ICA model in (\ref{eq:nonlinearICA}). In the limit of infinite data, iVAE recovers the true latent components $\bs z$ up to permutation and signed scaling under some mild conditions on the mixing function $\bs f$, the sufficient statistics $\bs T$ and the auxiliary variable $\bs u$ \cite{Khemakhem2020}. Here, we focus on Gaussian distributed latent variables, in which case the variances of the latent components are required to vary enough based on the auxiliary variable $\bs u$, and the mixing function $\bs f$ is required to have continuous partial derivatives.

iVAE consists mainly of the encoder $\bs g$, the decoder $\bs h$ and the auxiliary function $\bs w$. The encoder $\bs g(\bs x, \bs u) = (\bs g_{\bs \mu}(\bs x, \bs u)^\top, \bs g_{\bs \sigma}(\bs x, \bs u)^\top)^\top$ maps the observed data $(\bs x, \bs u)$ to mean and variance vectors $\bs \mu_{\bs z|\bs x, \bs u}$ and $\bs \sigma_{\bs z|\bs x, \bs u}$. The mean and the variance are used to sample a new latent representation $\bs z'$ by using a reparametrization trick \cite{kingma2013auto}, which is generating $\bs z'$ as $\bs z' = \bs \mu_{\bs z|\bs x, \bs u} + \bs \sigma_{\bs z|\bs x, \bs u}^\top \bs \epsilon$, where $\bs \epsilon \sim N(\bs 0, \bs I)$ and $\bs I$ is an identity matrix. The estimate of the unmixing transformation $\bs q$ is obtained as $\bs g_{\bs \mu}$. \red{This means that we obtain the latent components $\bs z$ as $\bs \mu_{\bs z|\bs x, \bs u}$, which is a mean of the variational distribution given by the encoder.} Meanwhile, the decoder $\bs h(\bs z)$ estimates the true mixing function $\bs f(\bs z)$ and maps the latent representation $\bs z'$ back to observable data $\bs x'$. The auxiliary function $\bs w(\bs u)$ estimates the parameters $\bs \lambda(\bs u)$ of (\ref{eq:sourcedist}) in the segment given by $\bs u$. The functions $\bs g, \bs h$ and $\bs w$ are deep neural networks, which are flexible function estimators and we denote their parameters as $\bs \theta_{\bs g}, \bs \theta_{\bs h}$, and $\bs \theta_{\bs w}$, respectively, which must be estimated. Coming back to the generating model as in (\ref{eq:generativemodel}), we now have that
\begin{align}
    p_{\bs \theta_{\bs h},\bs \theta_{\bs w}}(\bs x, \bs z|\bs u) = p_{\bs \theta_{\bs h}}(\bs x|\bs z)p_{\bs \theta_{\bs w}}(\bs z|\bs u),
\end{align}
where $p_{\bs \theta_{\bs h}}(\bs x|\bs z)$ is defined as in (\ref{eq:f_eps}). iVAE learns simultaneously the full latent generative model and the variational approximation $q_{\bs\theta_{\bs g}}(\bs z | \bs x, \bs u)$ of its true distribution $p_{\bs q}(\bs z | \bs x, \bs u)$. The parameters $\bs \theta = (\bs \theta_{\bs g}^\top, \bs \theta_{\bs h}^\top, \bs \theta_{\bs w}^\top)^\top$ are estimated by maximizing the lower bound of the data log-likelihood defined by
\begin{equation}
\begin{split}
    \mathcal{L}(\bs \theta | \bs x, \bs u) &\geq 
    E_{q_{\bs \theta_{\bs g}}(\bs z|\bs x,\bs u)}\big (\text{log}\, p_{\bs \theta_{\bs h}}(\bs x | \bs z)  + \text{log}\,p_{\bs \theta_{\bs w}}(\bs z | \bs u) - \text{log}\,q_{\bs \theta_{\bs g}}(\bs z | \bs x, \bs u) \big).
\end{split}
\label{eq:elbo_ivae}
\end{equation}
The variational approximation $q_{\bs \theta_{\bs g}}(\bs z | \bs x, \bs u)$ is obtained using a reparametrization trick. The schematic presentation of iVAE is illustrated in Fig.~\ref{fig:ivae}. In our simulation studies in Section~\ref{sec:simulations} we assume the source density distribution $p_{\bs\theta_{\bs w}}(\bs z | \bs u)$ to be Gaussian, where the mean and variance vectors, $\bs \mu_{\bs z|\bs u}$ and $\bs \sigma_{\bs z|\bs u}$, are given by the auxiliary function $\bs w(\bs u)$. Hence, in practice we have $p_{\bs\theta_{\bs w}}(\bs z | \bs u) = N(\bs z| \bs \mu_{\bs z|\bs u}, \text{diag}(\bs \sigma_{\bs z|\bs u}))$, $q_{\bs\theta_{\bs g}}(\bs z | \bs x, \bs u) = N(\bs z| \bs \mu_{\bs z|\bs x, \bs u}, \text{diag}(\bs \sigma_{\bs z|\bs x, \bs u}))$, and $p_{\bs\theta_{\bs h}}(\bs x | \bs z) = N(\bs x| \bs x', \beta \bs I)$, where $\beta > 0$ is a constant close to zero as $p_{\bs\theta_{\bs h}}$ should estimate the true distribution $p_{\bs f}$ with infinitesimal variance. We use $\beta = 0.01$.

\section{Interpreting the latent representation}
\label{sec:scaled_mashap}

In this section, we introduce the scaled MASHAP values and explain how they can be used to evaluate variable importance for model with multiple outputs.

MASHAP and SAGE, as defined in Section~\ref{sec:background_2}, estimate successfully the population level importance of input variables for a single output variable. However, for a model with multiple outputs it would be beneficial to be able to compare the population level importance values across the different outputs, as well as to obtain an importance value for each input variable with respect to all output variables. Here, we suggest a heuristic procedure to obtain \red{scaled MASHAP values, which are} comparable importance values between different output variables, and to calculate \red{average scaled MASHAP values, which are interpreted as} a population level importance values for the models with multiple output variables. \red{The procedure to obtain scaled MASHAP and average scaled MASHAP values is summarized in Algorithm~1, and discussed in detail next.

\begin{algorithm}
\red{
\caption{An algorithm for calculating the scaled MASHAP values and average scaled MASHAP values.}\label{alg:scaled_mashap}
\textbf{Input:} $n \times K$ data matrix $\bs X$, function $\tilde{\bs \gamma}^*: \mathbbm{R}^K \rightarrow \mathbbm{R}^H$\;
\textbf{Output:} Scaled MASHAP values $\bs V$; Average scaled MASHAP values $\bs v^*$\;
\textbf{Initialize:} $H \times K$ matrix $\bs V$; $1 \times K$ vector $\bs v^*$\;
\For{$i = 1$ \KwTo $H$}{
    Set $\tilde{\gamma} = \tilde{\gamma}^*_k$\;
    \For{$k = 1$ \KwTo $K$}{
        Calculate SHAP values $\tilde{a}_{k,1}, \dots \tilde{a}_{k,n}$ as in (\ref{eq:shap})\;
        Calculate MASHAP value $v_k$ as in (\ref{eq:mashap})\;
        Set $\bs V[i, k] \gets v_k$\;
    }
    Set $\bs V[i, ] \gets \frac{\bs V[i, ]}{\sum_{k=1}^K \bs V[i, k]}$\;
}
\For{$k = 1$ \KwTo $K$}{
    Set $\bs v^*[k] \gets \frac{\sum_{i=1}^H(\bs V[i,k])}{H}$
}
}
\end{algorithm}

Let $\bs \gamma^*$ be a model or a function with $H$-dimensional output. As SHAP and MASHAP assume a function with a single output, we calculate the MASHAP values independently for each output $i=1, \dots, H$ of the function $\bs \gamma^*$ by setting $\tilde{\gamma} = \gamma^*_i$. Then, we obtain the contributions of each input variable to each output varibale. However, by using this procedure, the MASHAP values are not comparable across different input variables. This is because the MASHAP values give only the relative importance of each input variable to a single output variable. As the distributions and scales of the output variables might differ, the sum of the MASHAP values of the input variables is not constant. Hence, the values are not relative to each other across the output variables, and comparisons of the MASHAP values might lead to false interpretations.}

To make the MASHAP values comparable, we suggest scaling the values to unit sum for each individual output variable. The scaling ensures that all output variables are weighted identically making the scaled MASHAP values comparable also across the outputs. In addition, we propose \red{average scaled MASHAP values} to identify the importance of the input variables for a model with multiple output variables. Let $\bs V$ be a matrix whose rows $\bs v^k$ contain the scaled MASHAP values for the $k$th \red{input} variable. By taking means column-wise, we get a $K$-dimensional vector $\bs v^*$ giving the relative importance of each input variable. 

\red{In context of nonlinear BSS, the scaled MASHAP values can be used to evaluate the contribution of each observed variable to a latent component, and the contribution of each latent component to an observable variable. By calculating the average scaled MASHAP values for the BSS model's mixing estimate, we obtain importance values for each latent component. When the average scaled MASHAP value is high, the latent component explains high proportion of the original variables, and when it is low, the latent component does not have much impact on most of the variables.}


The interpretations based on scaled MASHAP and average scaled MASHAP values are demonstrated \red{in more detail} in Section~\ref{sec:example}.

\section{Simulations}
\label{sec:simulations}

\red{While we have the theoretical results of identifiability when the variances of latent components are varying enough based on the auxiliary variable $\bs u$, these results apply only in limit of infinite data. In real life applications, with finite data, there is however no guarantee that the identifiablity conditions are fulfilled. In this section, we aim to study the performance of iVAE in various spatial scenarios with finite sample size using the simulation studies. We consider six different simulation settings, where the observed data are generated from nonlinear ICA model (\ref{eq:nonlinearICA}). The underlying latent components $\bs z$ are generated differently in each setting. Some of the settings exhibit nonstationary variance, meaning that the identifiability conditions are fulfilled, while some settings are stationary, for which the identifiability conditions are not fulfilled.} We compare iVAE to SBSS and SNSS, although they are developed for linear mixing and are thus not optimal when the mixing function is nonlinear. In addition, iVAE is compared against a modified version of time contrastive learning (TCL) \cite{HyvarinenMorioka2016}, where the auxiliary variable is a spatial segmentation instead of a time segmentation. TCL exploits nonstationarity in variance when solving the BSS problem and can estimate nonlinear unmixing transformations. The simulations can be reproduced using R 4.3.0 \cite{Rcoreteam} together with R packages SpatialBSS \cite{SpatialBSS}, gstat \cite{gstat} and NonlinearBSS. \red{The NonlinearBSS package, which is available at \url{https://github.com/mikasip/NonlinearBSS}, contains an R implementation of spatial iVAE and some methods to generate nonstationary spatial data from the nonlinear ICA model (\ref{eq:nonlinearICA}). The simulations were executed on the CSC Puhti cluster, a high-performance computing environment.}

\begin{figure}[htb]
\centerline{\includegraphics[width=0.55\textwidth]{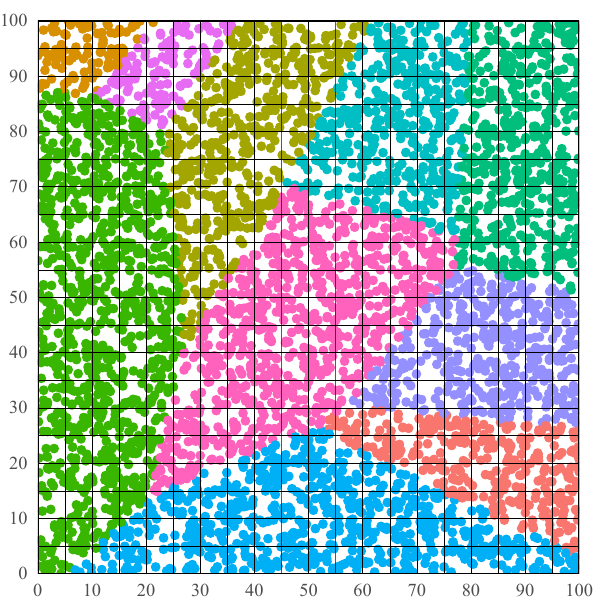}}
\caption{Five thousand simulated points in a $100 \times 100$ field based on ten randomly generated clusters. The colors indicate true cluster membership and the grid on top of the plot illustrates the spatial segmentation used by TCL and iVAE.}
\label{fig:clusters}
\end{figure}

In all simulation settings, $n=5000$ locations $\bs s_j$, $j=1,\dots,n$, are sampled uniformly in the spatial domain $\mathcal S  = [0, 100] \times [0, 100]$. For each location $\bs s_j$ a three-variate observation $\bs x_j = \bs x(\bs s_j)$ is generated, meaning that $d=3$ in all simulations. In Settings 1, 2 and 3, the data consist of ten clusters and are generated using different parameters in each cluster. The clusters are generated by sampling ten random points and dividing the field so that each cluster contains the locations that have the smallest distance to the corresponding point. Notice that we did not use the ground truth segmentation as an auxiliary variable in modeling as was done in previous studies, \red{such as \cite{HyvarinenMorioka2016, HyvarinenMorioka2017, Hyvarinen2019, Khemakhem2020}}, in a time series context. Instead, we used small enough segments to allow the model to approximate the ground truth clusters without having prior information on them. This is illustrated in Fig.~\ref{fig:clusters}, where $n=5000$ uniformly sampled points in ten clusters are presented. \red{In the figure, the colors represent the ten generated ground truth clusters and the grid on top of the plot illustrates the spatial segmentation used by iVAE and TCL.}

\textbf{Settings 1 and 2.} The latent field consists of identically and independently distributed (iid) three-variate Gaussian vectors. Each cluster, $k=1,\dots, 10$, has a unique, diagonal covariance matrix $\bs \Sigma_k = \text{diag}(\sigma_{1,k}, \sigma_{2,k}, \sigma_{3,k})$, where $\sigma_{1,k}, \sigma_{2,k}, \sigma_{3,k} \sim \text{Unif}(0.1,5)$. In Settings 1 and 2 the mean vectors are $\bs \mu_k = (0, 0, 0)^\top$ and $\bs \mu_k = (\mu_{1,k}, \mu_{2,k}, \mu_{3,k})^\top$, with $\mu_{1,k}, \mu_{2,k}, \mu_{3,k} \sim \text{Unif}(-5,5)$, respectively.

\textbf{Setting 3.} The latent field consists of three-variate Gaussian fields with a Matern correlation structure. Each cluster has its own parameters for the Matern covariance function
\begin{align}
    C_M(h; \nu, \phi)=\frac{1}{2^{\nu-1}\Gamma(\nu)}\left(\frac{h}{\phi}\right)^\nu K_\nu\left(\frac{h}{\phi}\right),
\end{align}
where $h = || \bs s  - \bs s' || $ is the Euclidean distance between two locations $\bs s $ and $\bs s'$, $\nu > 0$ is a scale parameter, $\phi > 0$ is a range parameter, $\Gamma$ is the gamma function, and $K_\nu$ is the modified Bessel function of the second kind with shape parameter $\nu$. The range and scale parameters are generated independently from $\nu_i \sim \text{Unif}(0.1, 5)$ and $\phi_i \sim \text{Unif}(0.5, 8)$.

\textbf{Settings 4 and 5.} The latent field is a single three-variate Gaussian random field with a Matern correlation structure. The Matern parameters for Setting 4 are $(\nu_1, \phi_1) = (0.5, 15), (\nu_2, \phi_2) = (2,20)$, and $(\nu_3, \phi_3)=(0.2, 10)$. In Setting 5, the parameters are $(\nu_1, \phi_1) = (1, 5), (\nu_2, \phi_2) = (2,3)$, and $(\nu_3, \phi_3)=(6, 2)$. Setting 4 shows strong spatial dependence and higher variability in variance, whereas Setting 5 shows weak spatial dependence and low variability in variance.

\textbf{Setting 6.} The latent field is a single three-variate random field with a nonstationary Matern correlation structure as presented in \cite{AnderesStein2011}:
\begin{equation}
\begin{split}
    &C_{M_{n}}(\bs s, \bs s'; \sigma, \nu, \phi) = \sigma(\bs s) \sigma(\bs s') r_1(\bs s)  r_1(\bs s')
    r_2(\bs s, \bs s')^{-1} \\
    &\left ( r_2(\bs s, \bs s')^{-1/2} h \right )^{\nu(\bs s,\bs s')}K_{\nu(\bs s,\bs s')}(r_2(\bs s, \bs s')^{-1/2} h),
\end{split}
\end{equation}
where $h=|| \bs s  - \bs s' ||$ and we denote $\nu(\bs s,\bs s')=(\nu(\bs s)+\nu(\bs s'))/2$,
\begin{equation}
\begin{split}
    &r_1(\bs s) = \left ( \frac{\phi^2(\bs s)/4 \nu(\bs s)}{\Gamma(\nu(\bs s))2^{\nu(\bs s) - 1}} \right )^{1/2} \quad \text{and} \\
    &r_2(\bs s, \bs s') = \left ( \frac{\phi^2(\bs s)}{8\nu(\bs s)} + \frac{\phi^2(\bs s')}{8\nu(\bs s')} \right ),
    \end{split}
\end{equation}
where $\sigma:\mathbb{R}^2 \rightarrow \mathbb{R}^+$, $\nu:\mathbb{R}^2 \rightarrow \mathbb{R}^+$ and $\phi:\mathbb{R}^2 \rightarrow \mathbb{R}^+$ are the local variance, shape, and range parameter functions, respectively. We choose these functions as
\begin{align}
    \sigma(\bs s; \bs d_\sigma, \alpha_\sigma) &= \text{log}( 1.1 + (\bs s^\top \bs d_\sigma)/\alpha_\sigma), \nonumber \\
    \nu(\bs s;\bs d_\nu, \alpha_\nu) &= (\bs s^\top \bs d_\nu)^{1/5}/\alpha_\nu + 0.1, \\
    \phi(\bs s; \bs d_\phi, \alpha_\phi, c_\phi) &= (\bs s^\top \bs d_\phi)/\alpha_\phi + c_\phi). \nonumber
\end{align}
For the first component $z_1$, we select the parameters $\bs d_\sigma = (1,1)$, $\alpha_\sigma = 2$, $\bs d_\nu = (0,1)$, $\alpha_\nu = 5$, $\bs d_\phi = (1,1)$, $\alpha_\phi = 5$, and $c_\phi = 10$; for $z_2$, we select $\bs d_\sigma = (0,1)$, $\alpha_\sigma = 1.5$, $\bs d_\nu = (1,0)$, $\alpha_\nu = 4$, $\bs d_\phi = (1,1)$, $\alpha_\phi = -8$, and $c_\phi = 40$; for $z_3$, we select $\bs d_\sigma = (1,0)$, $\alpha_\sigma = 2$, $\bs d_\nu = (1,1)$, $\alpha_\nu = 3$, $\bs d_\phi = (0,1)$, $\alpha_\phi = 4$, and $c_\phi = 10$. By using the above correlation structure, the variance, range, and scale parameters of the random field vary in space differently within each component.

Settings 1 and 2 are considered the \red{easiest} settings for the iVAE model as variances (and means in Setting 2) of the latent fields explicitly change between the clusters. These settings \red{are spatial variants of the time series settings of some previous simulation studies, such as \cite{Khemakhem2020, HyvarinenMorioka2016}, where the latent components have varying means/variances based on some time segements.} The settings aim to set a baseline performance under conditions, \red{where the variances are explicitly changing based on our chosen auxiliary variable}. Meanwhile, Settings 4 and 5 are used to measure the performance when the latent fields are stationary, \red{and thus not identifiable in theory}. Of these two, Setting 4 has higher variability in \red{sample} mean and in \red{sample} variance through the latent fields making it more optimal for iVAE. Settings 3 and 6 are chosen to illustrate the performance when the latent fields are nonstationary.

\red{\textbf{Mixing procedure.} The observed data $\bs x$ are generated by applying a mixing function $\bs f_L$ to the latent fields $\bs z$ as $\bs x = \bs f_L(\bs z)$.} The mixing function $\bs f_L$ is generated using multilayer perceptron (MLP) with $L$ layers following previous studies \cite{HyvarinenMorioka2016,Hyvarinen2019,Khemakhem2020}. In making the mixing function invertible and differentiable, the number of hidden units in each layer equals the number of latent components and the nonlinear activation functions are smooth bijective functions, such as a hyperbolic tangent or exponential linear unit (ELU) \cite{elu2015}. In addition, the mixing matrices of the layers of MLP are normalized to have unit length row and column vectors to guarantee that none of the latent components vanish in the mixing process. Let $\omega_i$ be the activation function of $i$th layer and $\bs B_i$ be the normalized mixing matrix of $i$th layer. Then, the mixing function $\bs f_L$ is defined as
\begin{align}
    \bs f_L(\bs z) = \begin{cases}
        \omega_L(\bs B_L \bs z),\quad L = 1, \\
        \omega_L(\bs B_L \bs f_{L-1}(\bs z)),\quad L \in \{2,3,\dots\}.
    \end{cases}
\end{align}
In these simulations, we used linear activation $\omega_L(x)=x$ for the last layer and ELU activation 
\begin{align}
\omega_i(x)=\begin{cases}  
    x,\quad x \geq 0, \\
    \text{exp}(x) - 1,\quad x < 0,
\end{cases}
\end{align} $i = 1,\dots, L-1$, for the other layers. This means that the mixing function $\bs f_1$ with one layer corresponds to a linear mixing case. The simulations are repeated 1000 times with 1, 2 and 3 mixing layers.

\textbf{Model specifications.} \red{In iVAE's encoder, decoder, and auxiliary function}, we used three hidden layers with 128 units \red{and leaky rectified linear unit (ReLU) activation to ensure that the model is capable of estimating the possible nonlinearities in mixing/unmixing functions and in the function $\bs \lambda(\bs u)$, which is modelled by the auxiliary function.} Likewise, in TCL, we used three hidden layers with 128 units and leaky ReLU activation. For both models, iVAE and TCL, a $20 \times 20$ segmentation was used as an auxiliary variable, meaning that the spatial domain was divided into $m=400$ equally sized squares. This segmentation is illustrated in Fig~\ref{fig:clusters}. iVAE and TCL were trained for 200 epochs with a batch size of 64. The initial learning rate was 0.01, which was decreased with second-order polynomial decay over 10000 training steps until the learning rate of 0.0001. SBSS and SNSS algorithms used two ring kernels defined by the parameters $\bs r_1 = (0,2)$ and $\bs r_2 = (2,4)$. Further, SNSS used a $10 \times 10$ segmentation. The sizes of the rings and the segmentation for SNSS yielded the best average performance in an initial simulation containing 100 trials of each setting with various segmentations and ring sizes.

\textbf{Performance index.} We used the mean correlation coefficient (MCC), which is a widely used performance metric in nonlinear ICA, to measure the performance of different methods, e.g. \cite{HyvarinenMorioka2016, HyvarinenMorioka2017, Khemakhem2020, Hyvarinen2019}. MCC is a function of the correlation matrix $\bs K = \text{Cor}(\hat{\bs z}, \bs z)$, where $\hat{\bs z}$ includes the estimated latent components, and is calculated as
\begin{align}
    \text{MCC}(\bs K)=\frac{1}{p} \sup_{\bs P  \in \mathcal{P}} \text{tr}(\bs P\, \abs(\bs K)),
\label{eq:MCC}
\end{align}
where $\mathcal{P}$ is a set of all possible permutation matrices, $\text{tr}(\cdot )$ is the trace of a matrix, and $\abs(\cdot)$ denotes taking the absolute value of a matrix elementwise. The optimal MCC value is one in which case the estimated and true sources are perfectly correlated up to their signs.

\begin{figure*}
\centerline{\includegraphics[width=0.99\textwidth]{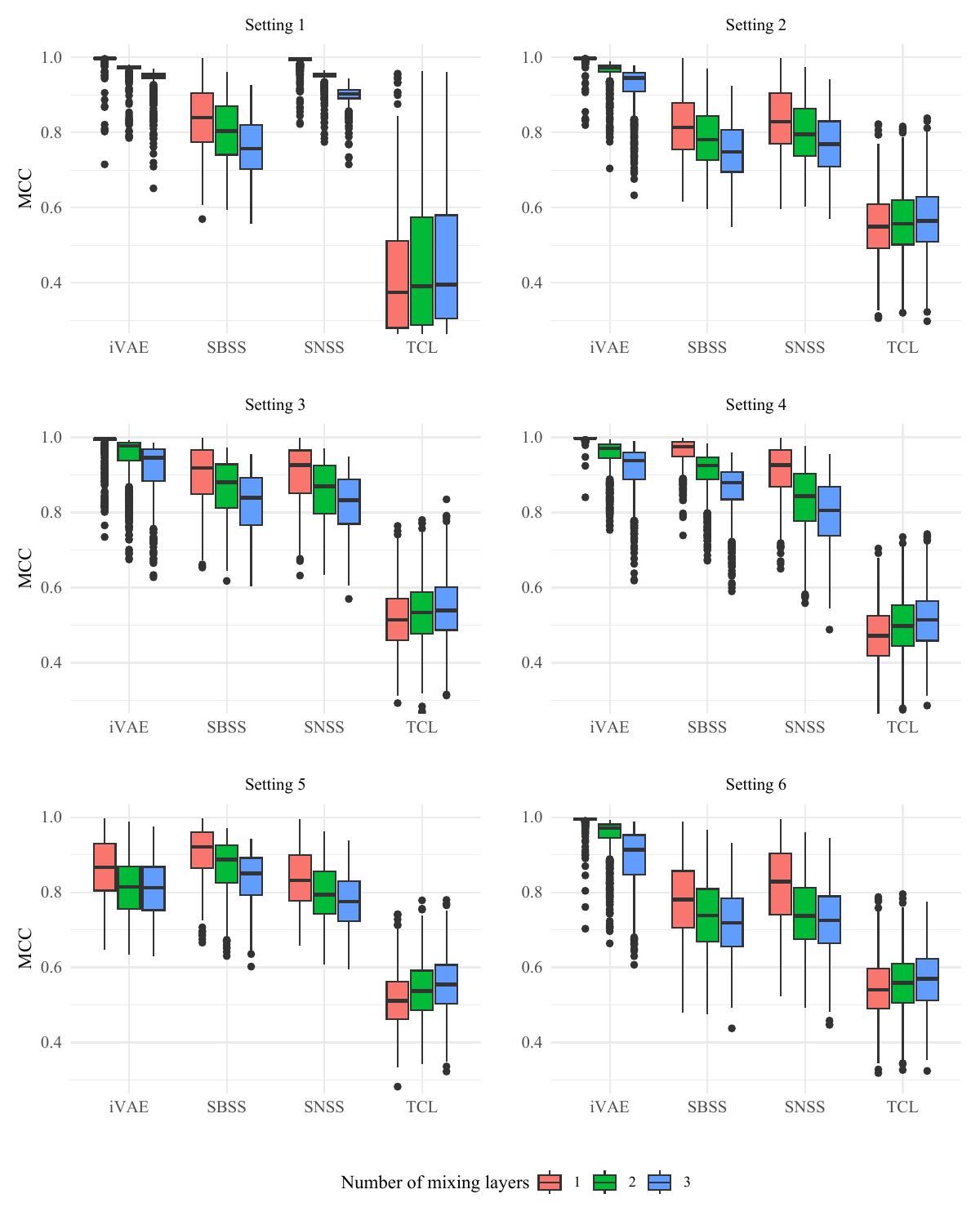}}
\caption{Boxlots of the mean correlation coefficients (MCC) of 1000 simulation trials of Settings 1-6 for iVAE, SBSS, SNSS and TCL. 
The color of the boxplots indicate the number of mixing layers where one layer corresponds to linear mixing.}
\label{fig:sim_all}
\end{figure*}

\textbf{Results.} The resulting MCC values are presented in Fig.~\ref{fig:sim_all}. 
In Settings 1, 2, 3 and 6, \red{where the variances of the latent fields were varying based on the spatial location, and thus fulfilling the identifiability conditions}, iVAE showed superior performance as compared with SBSS, SNSS, and TCL. In Setting~1, where the \red{latent fields} are zero-mean Gaussian with varying variances between the clusters, SNSS performed almost as well as iVAE, especially when the number of mixing layers was one. However, in Setting~2, where the mean also changed between the clusters, the performance of SNSS dropped considerably, whereas the performance of iVAE remained good. In Setting~3, SBSS and SNSS had similar performances, but the methods did not outperform iVAE, especially when the number of mixing layers was high. Meanwhile, Settings 4 and 5 are \red{stationary and thus}, in theory, more favorable for SBSS. \red{Surprisingly, in Setting~4, where the Matern covariance function's range parameters $\phi$ were high and the scale parameters $\nu$ were low, iVAE outperformed SBSS. This might be because such Matern parameters lead to high spatial dependence and larger variability in sample variance compared to Setting~5, which means that the mean and the variance are changing in some degree through out the spatial domain allowing the identifiability}. However, in Setting 5, where the \red{parameters were selected so that the} mean and the variance were more stable throughout the field, SBSS slightly outperformed iVAE. In Setting~6, iVAE was the only method that was reliably capable of separating the sources. \red{In conclusion, as long as the sample variance has enough variability in space, iVAE recovers the latent components well and outperform the competing methods.} 

\section{Real data example}
\label{sec:example}

In this section we demonstrate the spatial iVAE, SBSS and SNSS methods to a dataset derived from GEMAS geochemical mapping project \cite{reimann2014chemistry}, which is available in the R package robCompositions \cite{filzmoser2018applied}. \red{The application is two-fold. First, the latent fields are estimated with each method and the latent representations are compared using scaled MASHAP values. As we do not have any ground truth latent fields available as in simulations in Section~5, 
only the interpretability of the latent representations are compared. Second, we study if the prediction power can be increased by predicting the latent fields to new locations instead of predicting the observed variables directly. The predicted latent fields are then back transformed to the original variables by using the estimated mixing trasformations provided by iVAE, SBSS and SNSS.} \red{TCL is not applied for the data as it does not estimate the mixing transformation $\bs f$. For this reason, the scaled MASHAP values cannot be utilized for evaluating the importance of the latent components. For the same reason, the method is incapable of back transforming the estimated latent components, and thus cannot be applied for prediction purposes.} The code for the analysis of the data is available at \url{https://github.com/mikasip/NonlinearSNSS}. \red{The analysis was executed on a laptop running Windows 10, equipped with an Intel Core i5 processor (1.7 GHz) and 16 GB RAM.}

\begin{figure}
\centerline{\includegraphics[width=0.5\textwidth]{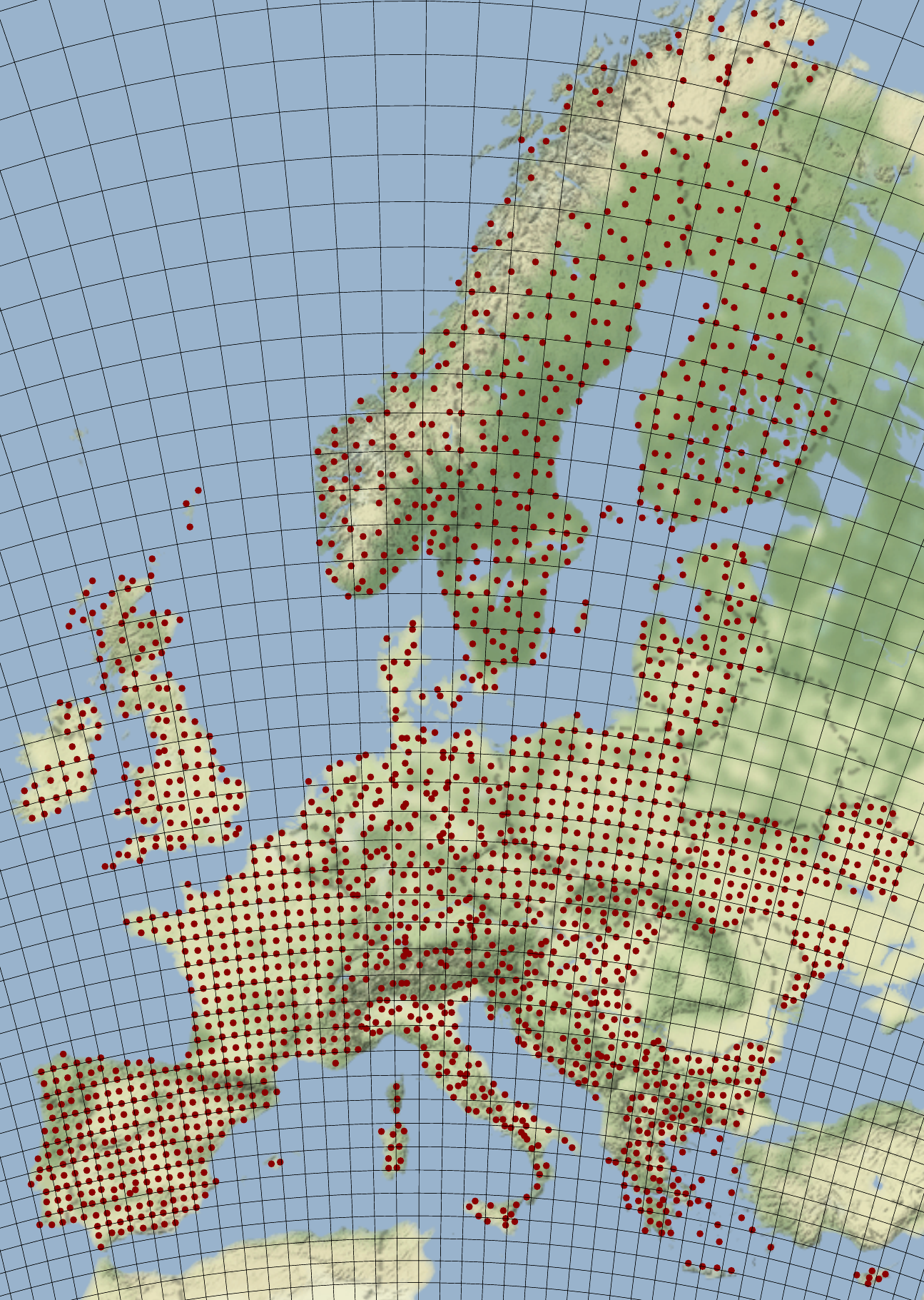}}
\caption{Sample locations of GEMAS dataset and 100km grid representing the segmentation used in iVAE model.}
\label{fig:points_with_grid}
\end{figure}

The dataset consists of 2108 consentration measurements of 18 chemical elements (Al, Ba, Ca, Cr, Fe, K, Mg, Mn, Na, Nb, P, Si, Sr, Ti, V, Y, Zn, Zr) in agricultural soil samples. We dropped one observation measured in Canary islands as it lacks neighboring observations. Therefore $n=2107$. The sample locations are presented in Fig.~\ref{fig:points_with_grid}.

In this example, we are dealing with compositional data, meaning that the variables carry relative information to other variables rather than absolute information. In this dataset, the measurements are chemical elements' relative proportions of the whole soil sample measured as milligrams per kilogram. 
\red{Due to the sum constraints, compositional data do not follow the Euclidean geometry and it is custom to transform such data prior analysis. \cite{NordhausenOjaFilzmoserReimann2015} argue that in the context of BSS, the isometric log-ratio (ilr) \cite{egozcue2003isometric} transformation is a natural choice yielding full rank data. While ilr coordinates are not easy to interpret, they have the additional benefit that they have a one-to-one relationship with the centralized log-ratio (clr) transformation. The clr transformation is popular among practitioners, because, although the coordinates are singular, they are interpretable. Accordingly, the data are processed using an ilr transformation which reduces the dimension of the dataset by one, and therefore the dimension of the preprocessed data is $17$. For interpretation purposes the ilr coordinates are transformed to clr coordinates and their interpretation goes similarly as in the case of log-transformed data, see \cite{egozcue2003isometric} for further details.}


In this application, we use the same iVAE model hyperparameters as in the simulation studies \red{of Section~\ref{sec:simulations}}. As an auxiliary variable, we use $10\text{km} \times 10\text{km}$ spatial segmentation, which is illustrated in Fig.~\ref{fig:points_with_grid}. Note that all spatial segments with zero observations are dropped. \red{For SBSS and SNSS, we select the parameters provided in \cite{MuehlmannBachocNordhausen2021}, where the same dataset was studied. Hence, we use one ring kernel with radius of 167 km for both SBSS and SNSS, and divide the spatial domain to four equally sized square segments for SNSS.}



\begin{figure}
\centerline{\includegraphics[width=0.4\textwidth]{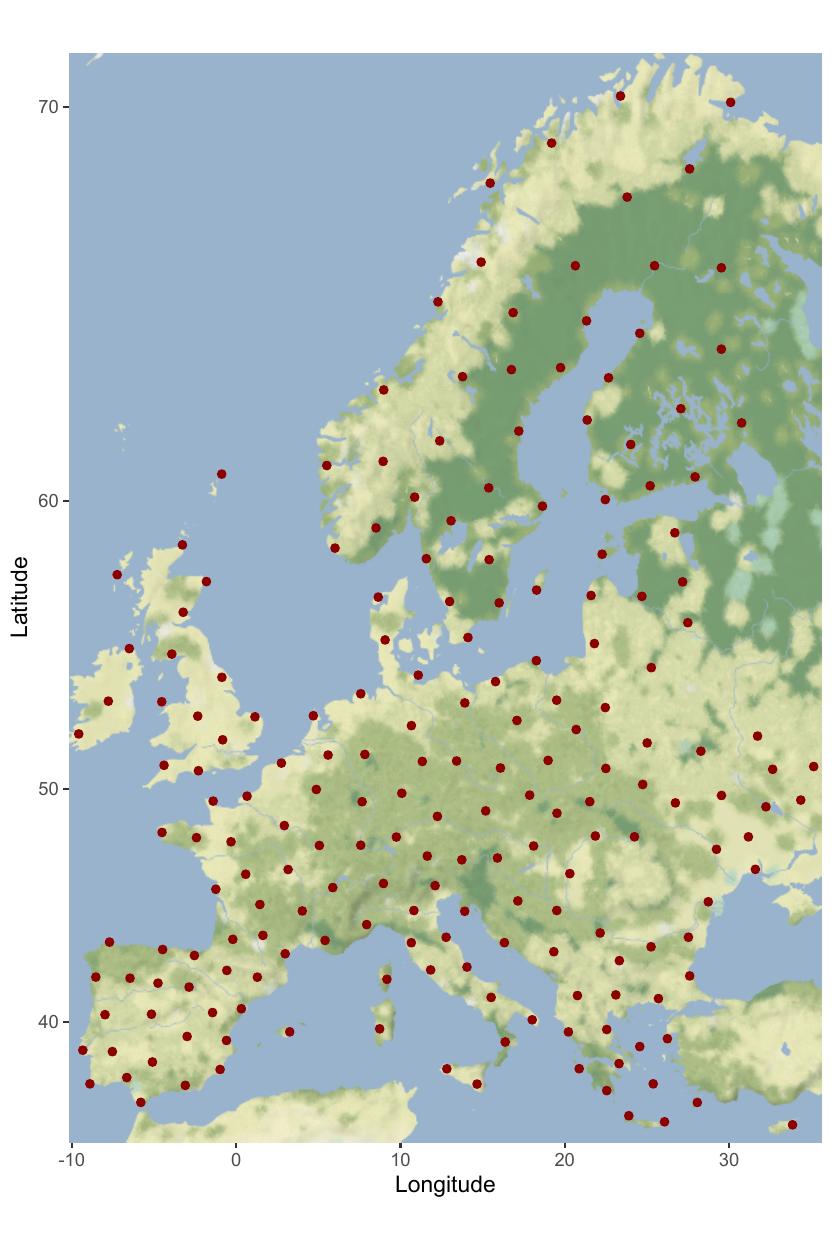}}
\caption{Sample locations used as background data for calculating the SHAP values.}
\label{fig:background_data}
\end{figure}

\red{First, the contributions of the latent components to the clr transformed variables are determined for iVAE, SBSS and SNSS by calculating the scaled MASHAP values for each method's mixing function estimate. For mixing function estimate, the input is the estimated latent components and the output is the ilr transformed observations, which are then transformed back in the clr transformed observations. Then, the scaled MASHAP values can be interpreted as importance of each latent feature to an original variable (in log scale). The averages of the scaled MASHAP values can be interpreted as importance of each latent feature to the whole dataset. The scaled MASHAP values are calculated by using 200 observations as background data. To ensure that the background data represents the whole dataset as well as possible, the background locations are selected one by one, so that the new location has the maximum distance to any previously selected location. The obtained background data are presented in Fig.~\ref{fig:background_data}.}

\begin{table}
\centering
\caption{The scaled MASHAP values for the decoder part of the trained iVAE model.}\label{table:SHAP_decoder}
\resizebox{\columnwidth}{!}{
\begin{tabular}{rrrrrrrrrrrrrrrrrr}
  \hline
 & IC1 & IC2 & IC3 & IC4 & IC5 & IC6 & IC7 & IC8 & IC9 & IC10 & IC11 & IC12 & IC13 & IC14 & IC15 & IC16 & IC17 \\
  \hline
clr(Al) & 0.099 & 0.119 & 0.154 & 0.169 & 0.093 & 0.075 & 0.039 & 0.041 & 0.023 & 0.080 & 0.049 & 0.025 & 0.008 & 0.007 & 0.007 & 0.007 & 0.007 \\
  clr(Ba) & 0.223 & 0.074 & 0.021 & 0.163 & 0.092 & 0.113 & 0.035 & 0.048 & 0.114 & 0.053 & 0.015 & 0.019 & 0.009 & 0.005 & 0.005 & 0.005 & 0.005 \\
  clr(Ca) & 0.046 & 0.410 & 0.016 & 0.096 & 0.175 & 0.051 & 0.046 & 0.059 & 0.030 & 0.018 & 0.009 & 0.023 & 0.005 & 0.004 & 0.004 & 0.004 & 0.004 \\
  clr(Cr) & 0.236 & 0.044 & 0.085 & 0.162 & 0.038 & 0.028 & 0.125 & 0.026 & 0.065 & 0.137 & 0.013 & 0.015 & 0.007 & 0.005 & 0.005 & 0.005 & 0.005 \\
  clr(Fe) & 0.404 & 0.098 & 0.101 & 0.060 & 0.032 & 0.025 & 0.034 & 0.037 & 0.026 & 0.045 & 0.076 & 0.028 & 0.009 & 0.006 & 0.006 & 0.006 & 0.006 \\
  clr(K) & 0.236 & 0.148 & 0.053 & 0.207 & 0.052 & 0.035 & 0.083 & 0.022 & 0.016 & 0.056 & 0.049 & 0.019 & 0.006 & 0.004 & 0.004 & 0.004 & 0.004 \\
  clr(Mg) & 0.197 & 0.223 & 0.139 & 0.086 & 0.090 & 0.021 & 0.033 & 0.089 & 0.034 & 0.045 & 0.012 & 0.014 & 0.004 & 0.003 & 0.003 & 0.003 & 0.003 \\
  clr(Mn) & 0.252 & 0.039 & 0.174 & 0.020 & 0.119 & 0.178 & 0.024 & 0.040 & 0.111 & 0.007 & 0.014 & 0.010 & 0.003 & 0.002 & 0.002 & 0.002 & 0.002 \\
  clr(Na) & 0.163 & 0.158 & 0.020 & 0.069 & 0.274 & 0.155 & 0.024 & 0.020 & 0.021 & 0.011 & 0.009 & 0.054 & 0.006 & 0.004 & 0.004 & 0.004 & 0.004 \\
  clr(Nb) & 0.055 & 0.338 & 0.083 & 0.037 & 0.031 & 0.036 & 0.097 & 0.095 & 0.020 & 0.036 & 0.097 & 0.045 & 0.007 & 0.006 & 0.006 & 0.006 & 0.006 \\
  clr(P) & 0.057 & 0.079 & 0.251 & 0.048 & 0.186 & 0.161 & 0.032 & 0.090 & 0.016 & 0.038 & 0.008 & 0.020 & 0.003 & 0.003 & 0.003 & 0.003 & 0.003 \\
  clr(Si) & 0.300 & 0.231 & 0.055 & 0.029 & 0.056 & 0.024 & 0.093 & 0.086 & 0.029 & 0.019 & 0.014 & 0.036 & 0.008 & 0.005 & 0.005 & 0.005 & 0.005 \\
  clr(Sr) & 0.175 & 0.366 & 0.058 & 0.046 & 0.033 & 0.016 & 0.102 & 0.019 & 0.035 & 0.020 & 0.018 & 0.092 & 0.005 & 0.004 & 0.004 & 0.004 & 0.004 \\
  clr(Ti) & 0.099 & 0.310 & 0.127 & 0.087 & 0.078 & 0.021 & 0.074 & 0.029 & 0.065 & 0.022 & 0.023 & 0.041 & 0.006 & 0.005 & 0.005 & 0.005 & 0.005 \\
  clr(V) & 0.381 & 0.081 & 0.100 & 0.082 & 0.026 & 0.022 & 0.043 & 0.017 & 0.085 & 0.105 & 0.014 & 0.022 & 0.005 & 0.005 & 0.005 & 0.005 & 0.005 \\
  clr(Y) & 0.057 & 0.290 & 0.087 & 0.038 & 0.087 & 0.047 & 0.076 & 0.073 & 0.116 & 0.026 & 0.058 & 0.011 & 0.013 & 0.005 & 0.005 & 0.005 & 0.005 \\
  clr(Zn) & 0.237 & 0.087 & 0.137 & 0.163 & 0.084 & 0.027 & 0.047 & 0.086 & 0.045 & 0.056 & 0.012 & 0.005 & 0.004 & 0.002 & 0.002 & 0.002 & 0.002 \\
  clr(Zr) & 0.241 & 0.309 & 0.026 & 0.106 & 0.041 & 0.045 & 0.044 & 0.021 & 0.021 & 0.052 & 0.040 & 0.025 & 0.010 & 0.005 & 0.005 & 0.005 & 0.005 \\
   \hline
 Average & \textbf{0.192} & \textbf{0.189} & 0.094 & 0.093 & 0.088 & 0.060 & 0.058 & 0.050 & 0.049 & 0.046 & 0.029 & 0.028 & 0.007 & 0.004 & 0.004 & 0.004 & 0.004 \\
   \hline
\end{tabular}
}
\end{table}

\red{For interpretation, we order latent components in decreasing order based on the average scaled MASHAP values. By inspecting the resulting scaled MASHAP values of iVAE's mixing function estimate provided in Table~\ref{table:SHAP_decoder}, we see that the latent components 1 and 2 are the most important ones as they have the highest average scaled MASHAP values of 0.192 and 0.189, whereas the values of other components are smaller than 0.094. The fact, that the average scaled MASHAP values of latent components IC13-IC17 are almost zero, indicates that these components are essentially noise and do not explain any of the chemical elements. Hence, fewer latent components would be enough to model the data. The average scaled MASHAP values of SBSS and SNSS, provided in~\ref{sec:appendix} in Table~\ref{table:sbss_shap} and Table~\ref{table:snss_shap}, are spread more evenly for the latent components. SBSS has average scaled MASHAP values from 0.035 to 0.114 and SNSS from 0.029 to 0.127. This indicates that none of the components can be dropped without losing information of the dataset. Also, based on the scaled MASHAP values, it is hard to determine the most interesting components, which makes the interpretation more difficult.}

\begin{figure}
  \begin{minipage}[b]{0.5\linewidth}
    \centering
    \includegraphics[width=.9\linewidth]{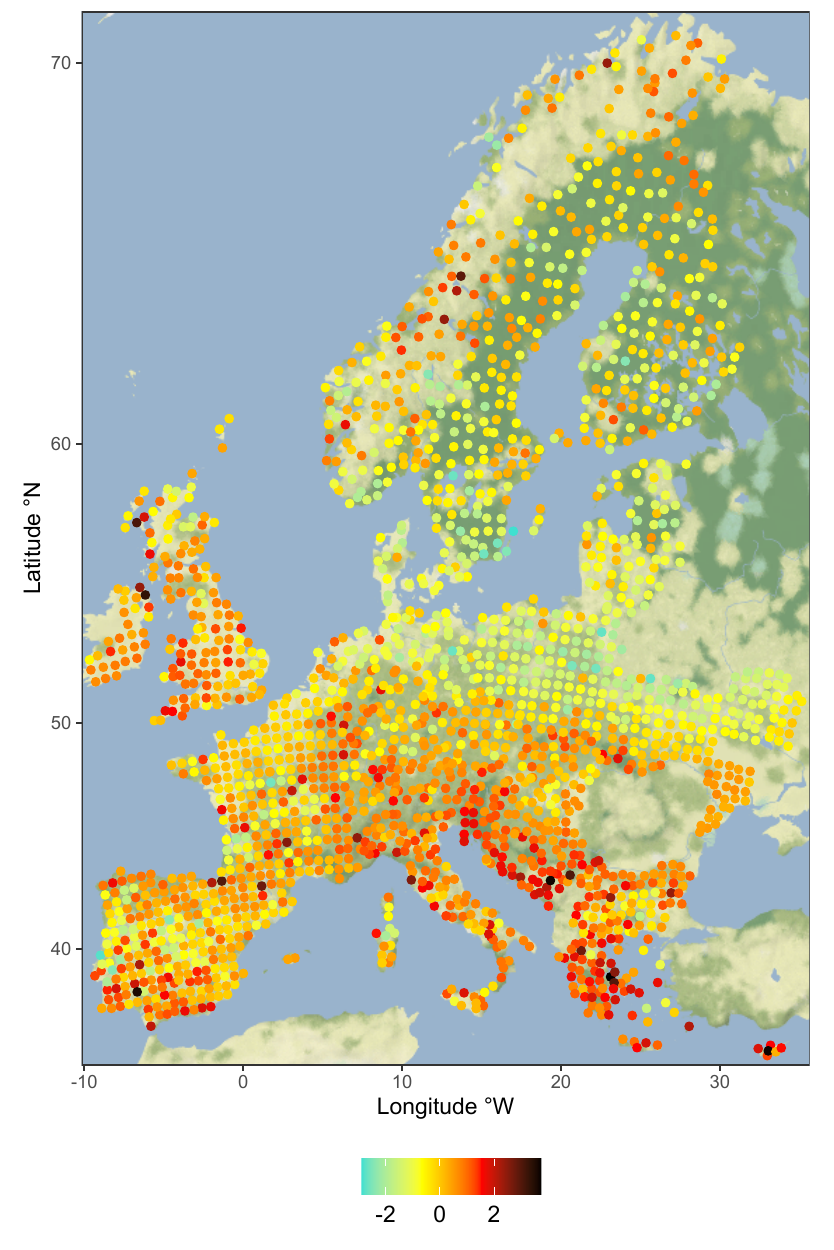} 
  \end{minipage}
  \begin{minipage}[b]{0.5\linewidth}
    \centering
    \includegraphics[width=.9\linewidth]{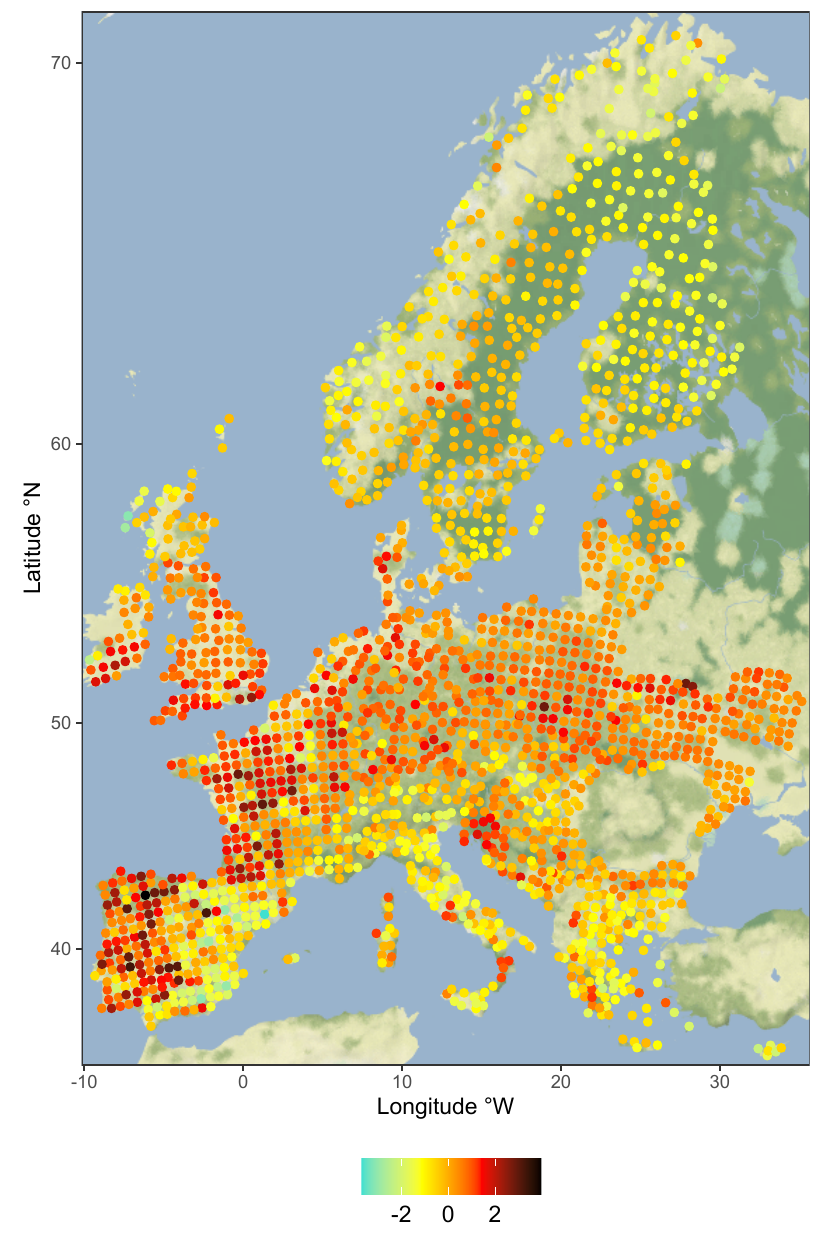} 
  \end{minipage}
\caption{The first (left) and the second (right) latent fields for GEMAS dataset recovered by iVAE.}
\label{fig:ics}
\end{figure}

\red{The two most important latent components provided by each of the methods are inspected more closely. The components based on iVAE are plotted in Figure~\ref{fig:ics}, and the components given by SBSS and SNSS are plotted in~\ref{sec:appendix} Figures~\ref{fig:ics_sbss} and~\ref{fig:ics_snss}, respectively. Of components recovered by iVAE, the first latent field shows north-south behaviour by having low values in north and higher ones in south, while the second latent field separates the middle Europe by having higher values there and lower ones in south and north. SBSS and SNSS both have one component separating the area from Ukraine to northern Germany (the second component for SBSS and the first for SNSS), while the other ones do not show that clear spatial structures. For SBSS, the first component has slightly lower values in diagonal area from United Kingdom to Greece while the rest of the map has high values, and the second component for SNSS shows any spatial behaviour only in area of Portugal and Spain.

We also calculated the scaled MASHAP values for iVAE's unmixing function estimate, where the original observations are treated as input together with the auxiliary variables, and the output is the latent components. By this, we can determine which observed variables contribute the most to the each of the latent components. Same background data is used as provided in Fig.~\ref{fig:background_data}. By looking at the scaled MASHAP values for the unmixing function estimate, provided in Table~\ref{table:SHAP_encoder}, it is evident that Ba and V contribute the most to the latent component 1, while the latent component 2 has highest contributions of Ca, Sr and Mg.}


\red{Finally, we study if the prediction power can be increased by predicting the latent components rather than the original observations.
We perform 10-fold crossvalidation by dividing the data randomly to 10 equally sized sets which are used one by one as a test data, while the rest of the data is used as a training data. The goal is to predict the observed chemical elements to the locations of the test data. As the data are multivariate and the variables are dependent on each other, the options are to use some multivariate modeling approach which takes the dependencies into account, or to find independent latent components which can be modelled individually. As a multivariate modeling approach we consider cokriging, which takes cross dependencies of each variable pair into account. Cokriging is compared against latent component approaches, where the latent components are found by iVAE, SBSS and SNSS, and the independent latent components are modelled individually using ordinary kriging and universal kriging. Ordinary kriging assumes a stationary mean in the space whereas universal kriging allows a trend in space. For details about kriging and cokriging, see e.g. \cite{chiles2018fifty, goovaerts1998ordinary}.

In cokriging, we fit Matern variogram models for all variograms and cross-variograms of the ilr transformed variables. For variogram models, we select a mutual range parameter by calculating the mean of range parameters of each variable's fitted univariate variograms. The mutual range parameter is selected in order to have a positive definite cokriging system, which is required to calculate predictions to new locations. The Matern parameters for each variogram and cross-variogram are fitted using ordinary least squares (OLS) method. As the predicting with cokriging is computationally very expensive with such many variables and large sample size, we use only 50 nearest points for cokriging predictions to make it feasible.

In ordinary and universal kriging, we fit Matern, exponential and spherical variogram models for each independent component and select the best based on the OLS method. For ordinary kriging, we allow the mean of the spatial field to be a linear function of spatial coordinates. To make a fair comparision, we use only 50 nearest points also for kriging predictions. The predicted latent components are transformed back to ilr variables using each method's mixing function estimate. At last, the predicted data are transformed from ilr space to clr space, where the performance is evaluated. As performance measures, we use mean squared error (MSE) mean absolute error (MAE) and root mean squared error (RMSE). MSE, MAE and RMSE are calculated as $MSE = \frac{1}{n} (x - \hat{x})^2$, $MAE = \frac{1}{n} |(x - \hat{x})|$ and $RMSE = \sqrt{\frac{1}{n} (x - \hat{x})^2}$, where $x$ is the true value and $\hat{x}$ is the estimated one. 
The average performances over the 10-fold crossvalidation are presented in Table~\ref{table:pred_errors}.} The lowest MSE, MAE and RMSE are obtained when iVAE is combined with ordinary kriging making it the best performing prediction method. SNSS combined with kriging slightly outperforms cokriging, and SBSS combined with kriging has very similar performance as cokriging. Universal kriging did not increase the performance, which indicates that the fields do not have linear trend in space. Because of the performance gain of iVAE combined with kriging, and the fact univariate kriging is computationally much more feasible than cokriging, we consider iVAE combined with kriging the best prediction method.

\begin{table}[ht]
\centering
\red{
\caption{MSE, MAE and RMSE prediction errors of 10-fold crossvalidation, where the chemical elements of the GEMAS dataset were predicted to new spatial locations using either cokriging or iVAE, SBSS and SNSS combined with ordinary or universal kriging.}
\label{table:pred_errors}}
\begin{tabular}{rrrrr}
  \hline
 Method & MSE & MAE & RMSE \\
  \hline
  Cokriging & 0.1811 & 0.2787 & 0.4252 \\
  iVAE + Ordinary kriging & \textbf{0.1768} & \textbf{0.2717} & \textbf{0.4201} \\
  iVAE + Universal kriging & 0.1770 & 0.2719 & 0.4203 \\
  SBSS + Ordinary kriging & 0.1819 & 0.2777 & 0.4261 \\
  SBSS + Universal kriging & 0.1825 & 0.2781 & 0.4268 \\
  SNSS + Ordinary kriging & 0.1792 & 0.2758 & 0.4230 \\
  SNSS + Universal kriging & 0.1797 & 0.2762 & 0.4236 \\
   \hline
\end{tabular}
\end{table}

\section{Conclusions}
\label{sec:conclusions}
In this paper we extended iVAE, first proposed by \cite{Khemakhem2020}, to the nonlinear SNSS setting. We discussed the theoretical background of the method and provided useful tools for interpreting the latent components through nonlinear mixing environment using SHAP values. We used simulation studies to illustrate the performance of iVAE method in nonlinear SBSS and SNSS settings, and applied the method for a geochemical dataset.

Simulation studies revealed that iVAE outperformed its competitors in all settings where the variability in variance of the latent fields was relatively high, that is, the variances changed relatively much between different locations. The variability was achieved either by having multiple (in our simulations, ten) clusters with varying variances, by having latent fields with a nonstationary correlation structure or by having latent fields with stationary correlation structure that yields strong spatial dependence and enough variability in sample variances in space. In the stationary setting, where the mean and variance did not change enough, SBSS still performed 
better.

In the geochemical application, we utilized iVAE, SBSS and SNSS to find the latent fields for the geochemical dataset. We interpreted the latent fields provided by SBSS, SNSS and iVAE by calculating the scaled MASHAP values for the mixing function estimates, and discovered which fields are the most important ones for each method. Based on the scaled MASHAP values, iVAE provided the easiest interpretable components, two of which were the most important ones. 
We examined the results further by comparing the two most important component of SBSS, SNSS and iVAE by plotting the corresponding latent fields. iVAE provided clearer structure for the components than the competing methods. In depth interpretations of the latent components, however, requires subject expertise and is beyond the scope of this paper. \red{We also studied if the predicition power can be increased by pre-processing the data using iVAE, and then predicting the independent latent components univariately, and backtransforming the predictions to the original variable space. The results suggest that by using iVAE as pre-processing method, the prediction power can be increased compared to alternative multivariate predicting approach.}

\section{Discussion}
\label{sec:discussion}

Based on the results of this paper, iVAE is a preferable method in settings, where the variances of the latent fields are not stable across space. However, in stationary settings where the sample mean and sample variance did not change enough, SBSS still performed better. In practice, this means for example having small range and large shape parameters in a Matern covariance function. However, many real-world spatial phenomena, such as temperature or humidity, often show high spatial dependence making iVAE the preferable method. To overcome the performance drop in stationary settings with low spatial dependence, it requires development of new models/methods which do not assume nonstationarity for identifiability. This is left for future work.

In the geochemical application, we calculated the average scaled MASHAP values of the latent components of spatial iVAE's mixing function estimate, and discovered that fewer components might be sufficient to model the dataset. iVAE allows modelling the data with fewer latent components than the observed varibles but currently there are, to the best of our knowledge, no methods for estimating the latent dimension. Hence, in future work, procedures for estimating the dimension of the latent representation will be developed.

In our simulation studies, TCL performed poorly in all settings. This is most likely due to nonoptimal simulation setups for the method. TCL relies on classifying observed data into chosen segments; when the classification task is too easy (e.g. we have data with changing mean or the number of real clusters is too low), TCL succeeds in classification without finding ICs. This problem was previously acknowledged also in \cite{Khemakhem2020}. Thus, various methods should be also compared in more favourable settings for TCL in future work.

In this paper, we only considered Gaussian fields and assumed Gaussian latent components in model specification. In the future, the effect of different source density distributions and the robustness properties under outliers or model mismatch must be studied, as was done for SBSS in \cite{robustSBSS}. Considering the flexibility of iVAE \red{in terms of possibilities for auxiliary data}, the method could be extended to various situations, such as for spatio-temporal data, as was also recently considered in a linear BSS framework \cite{MuehlmannDeIacoNordhausen2022}. For the latent field interpretation, we considered MASHAP based procedure for population level interpretations. In future work, other approaches, such as Shapley additive global importance \cite{sage} should be explored for latent component interpretation.

\section*{Acknowledgements}
We thank the three anonymous reviewers for helping
us improve and clarify this manuscript by their insightful comments. This work was partly supported by the Austrian Science Fund (P31881-N32), the Research Council of Finland (453691), the HiTEc COST Action (CA21163), the Vilho, Yrjö and Kalle Väisälä Foundation, and the Kone foundation. We also wish to acknowledge CSC – IT Center for Science, Finland, for computational resources.

\bibliographystyle{elsarticle-num} 
\bibliography{bibfile}

\begin{thebibliography}{10}
\expandafter\ifx\csname url\endcsname\relax
  \def\url#1{\texttt{#1}}\fi
\expandafter\ifx\csname urlprefix\endcsname\relax\def\urlprefix{URL }\fi
\expandafter\ifx\csname href\endcsname\relax
  \def\href#1#2{#2} \def\path#1{#1}\fi

\bibitem{GentonKleiber2015}
M.~G. Genton, W.~Kleiber, Cross-covariance functions for multivariate
  geostatistics, Statistical Science 30 (2015) 147--163.
\newblock \href {https://doi.org/https://doi.org/10.1214/14-STS487}
  {\path{doi:https://doi.org/10.1214/14-STS487}}.

\bibitem{MuehlmannNordhausenYi2021}
C.~Muehlmann, K.~Nordhausen, M.~Yi, On cokriging, neural networks, and spatial
  blind source separation for multivariate spatial prediction, IEEE Geoscience
  and Remote Sensing Letters 18~(11) (2021) 1931--1935.
\newblock \href {https://doi.org/10.1109/LGRS.2020.3011549}
  {\path{doi:10.1109/LGRS.2020.3011549}}.

\bibitem{comon2010handbook}
P.~Comon, C.~Jutten, Handbook of Blind Source Separation: {I}ndependent
  Component Analysis and Applications, Academic Press, 2010.
\newblock \href {https://doi.org/10.1016/C2009-0-19334-0}
  {\path{doi:10.1016/C2009-0-19334-0}}.

\bibitem{NordhausenOjaFilzmoserReimann2015}
K.~Nordhausen, H.~Oja, P.~Filzmoser, C.~Reimann, Blind source separation for
  spatial compositional data, Mathematical Geosciences 47~(7) (2015) 753--770.
\newblock \href {https://doi.org/10.1007/s11004-014-9559-5}
  {\path{doi:10.1007/s11004-014-9559-5}}.

\bibitem{bachoc2018spatial}
F.~Bachoc, M.~G. Genton, K.~Nordhausen, A.~Ruiz-Gazen, J.~Virta, Spatial blind
  source separation, Biometrika 107 (2020) 627--646.
\newblock \href {https://doi.org/10.1093/biomet/asz079}
  {\path{doi:10.1093/biomet/asz079}}.

\bibitem{MuehlmannBachocNordhausenYi2024}
C.~Muehlmann, F.~Bachoc, K.~Nordhausen, M.~Yi, Test of the latent dimension of
  a spatial blind source separation model, To appear in Statistica Sinica
  (2024).
\newblock \href {https://doi.org/10.5705/ss.202021.0326}
  {\path{doi:10.5705/ss.202021.0326}}.

\bibitem{MuehlmannBachocNordhausen2021}
C.~Muehlmann, F.~Bachoc, K.~Nordhausen, Blind source separation for
  non-stationary random fields, Spatial Statistics 47 (2022) 100574.
\newblock \href {https://doi.org/https://doi.org/10.1016/j.spasta.2021.100574}
  {\path{doi:https://doi.org/10.1016/j.spasta.2021.100574}}.

\bibitem{Hyvarinenetal:2023}
A.~Hyvärinen, I.~Khemakhem, H.~Morioka,
  \href{https://www.sciencedirect.com/science/article/pii/S2666389923002234}{Nonlinear
  independent component analysis for principled disentanglement in unsupervised
  deep learning}, Patterns 4~(10) (2023) 100844.
\newblock \href {https://doi.org/https://doi.org/10.1016/j.patter.2023.100844}
  {\path{doi:https://doi.org/10.1016/j.patter.2023.100844}}.
\newline\urlprefix\url{https://www.sciencedirect.com/science/article/pii/S2666389923002234}

\bibitem{hyvarinen1999nonlinear}
A.~Hyv{\"a}rinen, P.~Pajunen, {Nonlinear independent component analysis:
  Existence and uniqueness results}, Neural Networks 12~(3) (1999) 429--439.
\newblock \href {https://doi.org/https://doi.org/10.1016/S0893-6080(98)00140-3}
  {\path{doi:https://doi.org/10.1016/S0893-6080(98)00140-3}}.

\bibitem{kingma2013auto}
D.~P. Kingma, M.~Welling, {Auto-encoding Variational Bayes}, arXiv preprint
  arXiv:1312.6114 (2013).
\newblock \href {https://doi.org/https://doi.org/10.48550/arXiv.1312.6114}
  {\path{doi:https://doi.org/10.48550/arXiv.1312.6114}}.

\bibitem{higgins2017betavae}
I.~Higgins, L.~Matthey, A.~Pal, C.~Burgess, X.~Glorot, M.~Botvinick,
  S.~Mohamed, A.~Lerchner, {{$\beta$}-VAE: Learning basic visual concepts with
  a constrained variational framework}, in: International Conference on
  Learning Representations, 2017.

\bibitem{zhao2017infovae}
S.~Zhao, J.~Song, S.~Ermon, {InfoVAE: Information maximizing variational
  autoencoders}, arXiv preprint arXiv:1706.02262 (2017).
\newblock \href {https://doi.org/https://doi.org/10.48550/arXiv.1706.02262}
  {\path{doi:https://doi.org/10.48550/arXiv.1706.02262}}.

\bibitem{Khemakhem2020}
I.~Khemakhem, D.~Kingma, R.~Monti, A.~Hyv\"arinen, {Variational autoencoders
  and nonlinear {ICA}: A unifying framework}, in: International Conference on
  Artificial Intelligence and Statistics, PMLR, 2020, pp. 2207--2217.
\newblock \href {https://doi.org/https://doi.org/10.48550/arXiv.1907.04809}
  {\path{doi:https://doi.org/10.48550/arXiv.1907.04809}}.

\bibitem{halva2021disentangling}
H.~H{\"a}lv{\"a}, S.~Le~Corff, L.~Leh{\'e}ricy, J.~So, Y.~Zhu, E.~Gassiat,
  A.~Hyv{\"a}rinen, {Disentangling identifiable features from noisy data with
  structured nonlinear ICA}, Advances in Neural Information Processing Systems
  34 (2021) 1624--1633.

\bibitem{kernelshap}
S.~M. Lundberg, S.-I. Lee, A unified approach to interpreting model
  predictions, Advances in Neural Information Processing Systems 30 (2017).
\newblock \href {https://doi.org/https://doi.org/10.48550/arXiv.1705.07874}
  {\path{doi:https://doi.org/10.48550/arXiv.1705.07874}}.

\bibitem{HyvarinenMorioka2016}
A.~Hyv\"arinen, H.~Morioka, Unsupervised feature extraction by time-contrastive
  learning and nonlinear {ICA}, {Advances in Neural Information Processing
  Systems} 29 (2016).
\newblock \href {https://doi.org/https://doi.org/10.48550/arXiv.1605.06336}
  {\path{doi:https://doi.org/10.48550/arXiv.1605.06336}}.

\bibitem{HyvarinenMorioka2017}
A.~Hyv\"arinen, H.~Morioka, {Nonlinear {ICA} of temporally dependent stationary
  sources}, in: Artificial Intelligence and Statistics, PMLR, 2017, pp.
  460--469.

\bibitem{Hyvarinen2019}
A.~Hyv\"arinen, H.~Sasaki, R.~Turner, {Nonlinear {ICA} using auxiliary
  variables and generalized contrastive learning}, in: The 22nd International
  Conference on Artificial Intelligence and Statistics, PMLR, 2019, pp.
  859--868.
\newblock \href {https://doi.org/https://doi.org/10.48550/arXiv.1805.08651}
  {\path{doi:https://doi.org/10.48550/arXiv.1805.08651}}.

\bibitem{HalvaHyvarinen2020}
H.~H{\"a}lv{\"a}, A.~Hyv\"arinen, {Hidden Markov nonlinear {ICA}: Unsupervised
  learning from nonstationary time series}, in: Conference on Uncertainty in
  Artificial Intelligence, PMLR, 2020, pp. 939--948.
\newblock \href {https://doi.org/https://doi.org/10.48550/arXiv.1605.06336}
  {\path{doi:https://doi.org/10.48550/arXiv.1605.06336}}.

\bibitem{shapley1953value}
L.~S. Shapley, {17. A value for n-person games}, Princeton University Press,
  Princeton, 1953, pp. 307--318.
\newblock \href {https://doi.org/10.1515/9781400881970-018}
  {\path{doi:10.1515/9781400881970-018}}.

\bibitem{LIME2016}
M.~T. Ribeiro, S.~Singh, C.~Guestrin, {``Why should I trust you?'': Explaining
  the predictions of any classifier}, in: Proceedings of the 22nd ACM SIGKDD
  international conference on knowledge discovery and data mining, 2016, pp.
  1135--1144.
\newblock \href {https://doi.org/https://doi.org/10.1145/2939672.2939778}
  {\path{doi:https://doi.org/10.1145/2939672.2939778}}.

\bibitem{marcilio2020explanations}
W.~E. Marc{\'\i}lio, D.~M. Eler, {From explanations to feature selection:
  assessing SHAP values as feature selection mechanism}, in: 2020 33rd SIBGRAPI
  conference on Graphics, Patterns and Images (SIBGRAPI), Ieee, 2020, pp.
  340--347.
\newblock \href {https://doi.org/10.1109/SIBGRAPI51738.2020.00053}
  {\path{doi:10.1109/SIBGRAPI51738.2020.00053}}.

\bibitem{sage}
I.~Covert, S.~M. Lundberg, S.-I. Lee, Understanding global feature
  contributions with additive importance measures, Advances in Neural
  Information Processing Systems 33 (2020) 17212--17223.

\bibitem{Rcoreteam}
{R Core Team}, \href{https://www.R-project.org/}{R: A Language and Environment
  for Statistical Computing}, R Foundation for Statistical Computing, Vienna,
  Austria (2023).
\newline\urlprefix\url{https://www.R-project.org/}

\bibitem{SpatialBSS}
C.~Muehlmann, M.~Sipil\"a, K.~Nordhausen, S.~Taskinen, J.~Virta,
  \href{https://CRAN.R-project.org/package=SpatialBSS}{\texttt{SpatialBSS}:
  Blind Source Separation for Multivariate Spatial Data}, {\textsf{R}} package
  version 0.13-0 (2022).
\newline\urlprefix\url{https://CRAN.R-project.org/package=SpatialBSS}

\bibitem{gstat}
E.~J. Pebesma, Multivariable geostatistics in {S}: the gstat package, Computers
  \& Geosciences 30 (2004) 683--691.

\bibitem{AnderesStein2011}
E.~B. Anderes, M.~L. Stein, Local likelihood estimation for nonstationary
  random fields, Journal of Multivariate Analysis 102~(3) (2011) 506--520.
\newblock \href {https://doi.org/https://doi.org/10.1016/j.jmva.2010.10.010}
  {\path{doi:https://doi.org/10.1016/j.jmva.2010.10.010}}.

\bibitem{elu2015}
D.-A. Clevert, T.~Unterthiner, S.~Hochreiter, {Fast and accurate deep network
  learning by exponential linear units (ELUs)}, arXiv preprint arXiv:1511.07289
  (2015).
\newblock \href {https://doi.org/https://doi.org/10.48550/arXiv.1511.07289}
  {\path{doi:https://doi.org/10.48550/arXiv.1511.07289}}.

\bibitem{reimann2014chemistry}
C.~Reimann, M.~Birke, A.~Demetriades, P.~Filzmoser, P.~O'Connor, {Chemistry of
  Europe's agricultural soils, part A}, Schweizerbart'sche Verlagsbuchhandlung,
  Stuttgart, Germany, 2014.

\bibitem{filzmoser2018applied}
P.~Filzmoser, K.~Hron, M.~Templ, Applied compositional data analysis, Cham:
  Springer (2018).

\bibitem{egozcue2003isometric}
J.~J. Egozcue, V.~Pawlowsky-Glahn, G.~Mateu-Figueras, C.~Barcelo-Vidal,
  Isometric logratio transformations for compositional data analysis,
  Mathematical Geology 35~(3) (2003) 279--300.
\newblock \href {https://doi.org/https://doi.org/10.1023/A:1023818214614}
  {\path{doi:https://doi.org/10.1023/A:1023818214614}}.

\bibitem{chiles2018fifty}
J.-P. Chil{\`e}s, N.~Desassis, Fifty years of kriging, Handbook of mathematical
  geosciences: Fifty years of IAMG (2018) 589--612.

\bibitem{goovaerts1998ordinary}
P.~Goovaerts, Ordinary cokriging revisited, Mathematical Geology 30 (1998)
  21--42.

\bibitem{robustSBSS}
M.~Sipil{\"a}, C.~Muehlmann, K.~Nordhausen, S.~Taskinen, Robust second-order
  stationary spatial blind source separation using generalized sign matrices,
  Spatial Statistics 59 (2024) 100803.
\newblock \href {https://doi.org/https://doi.org/10.1016/j.spasta.2023.100803}
  {\path{doi:https://doi.org/10.1016/j.spasta.2023.100803}}.

\bibitem{MuehlmannDeIacoNordhausen2022}
C.~Muehlmann, S.~De~Iaco, K.~Nordhausen, Blind recovery of sources for
  multivariate space-time random fields, Stochastic Environmental Research and
  Risk Assessment 37 (2023) 1593--1613.
\newblock \href {https://doi.org/10.1007/s00477-022-02348-2}
  {\path{doi:10.1007/s00477-022-02348-2}}.

\end{thebibliography}

\appendix

\section{Additional results for a real data example}
\label{sec:appendix}

\begin{table}[ht]
\centering
\red{
\caption{Scaled MASHAP values for SBSS method's mixing function estimate.}\label{table:sbss_shap}
\resizebox{\columnwidth}{!}{
\begin{tabular}{rrrrrrrrrrrrrrrrrr}
  \hline
 & IC3 & IC2 & IC4 & IC1 & IC5 & IC6 & IC7 & IC12 & IC11 & IC15 & IC8 & IC17 & IC10 & IC9 & IC13 & IC16 & IC14 \\
  \hline
clr(Al) & 0.033 & 0.157 & 0.168 & 0.032 & 0.125 & 0.013 & 0.001 & 0.034 & 0.080 & 0.039 & 0.113 & 0.060 & 0.002 & 0.050 & 0.024 & 0.067 & 0.002 \\
  clr(Ba) & 0.193 & 0.074 & 0.068 & 0.063 & 0.050 & 0.054 & 0.013 & 0.079 & 0.040 & 0.033 & 0.133 & 0.065 & 0.028 & 0.046 & 0.014 & 0.013 & 0.032 \\
  clr(Ca) & 0.015 & 0.134 & 0.186 & 0.005 & 0.136 & 0.016 & 0.052 & 0.090 & 0.055 & 0.042 & 0.078 & 0.020 & 0.061 & 0.008 & 0.057 & 0.018 & 0.028 \\
  clr(Cr) & 0.310 & 0.016 & 0.147 & 0.038 & 0.069 & 0.081 & 0.045 & 0.031 & 0.037 & 0.006 & 0.007 & 0.012 & 0.005 & 0.044 & 0.076 & 0.022 & 0.053 \\
  clr(Fe) & 0.129 & 0.160 & 0.053 & 0.037 & 0.003 & 0.088 & 0.033 & 0.104 & 0.086 & 0.015 & 0.004 & 0.117 & 0.020 & 0.051 & 0.075 & 0.002 & 0.023 \\
  clr(K) & 0.170 & 0.047 & 0.102 & 0.058 & 0.089 & 0.121 & 0.085 & 0.015 & 0.064 & 0.013 & 0.084 & 0.091 & 0.003 & 0.010 & 0.009 & 0.025 & 0.013 \\
  clr(Mg) & 0.116 & 0.205 & 0.091 & 0.089 & 0.022 & 0.022 & 0.061 & 0.013 & 0.037 & 0.037 & 0.048 & 0.011 & 0.080 & 0.133 & 0.012 & 0.014 & 0.009 \\
  clr(Mn) & 0.160 & 0.038 & 0.046 & 0.029 & 0.025 & 0.015 & 0.012 & 0.043 & 0.011 & 0.135 & 0.053 & 0.131 & 0.098 & 0.013 & 0.049 & 0.033 & 0.109 \\
  clr(Na) & 0.073 & 0.075 & 0.054 & 0.297 & 0.086 & 0.103 & 0.010 & 0.015 & 0.026 & 0.038 & 0.009 & 0.050 & 0.017 & 0.037 & 0.004 & 0.055 & 0.050 \\
  clr(Nb) & 0.009 & 0.052 & 0.132 & 0.153 & 0.016 & 0.037 & 0.053 & 0.061 & 0.028 & 0.081 & 0.065 & 0.037 & 0.024 & 0.079 & 0.097 & 0.034 & 0.041 \\
  clr(P) & 0.065 & 0.100 & 0.094 & 0.029 & 0.004 & 0.111 & 0.019 & 0.112 & 0.035 & 0.070 & 0.027 & 0.029 & 0.067 & 0.065 & 0.049 & 0.088 & 0.035 \\
  clr(Si) & 0.152 & 0.228 & 0.003 & 0.138 & 0.154 & 0.027 & 0.032 & 0.019 & 0.021 & 0.039 & 0.016 & 0.018 & 0.011 & 0.094 & 0.030 & 0.003 & 0.016 \\
  clr(Sr) & 0.113 & 0.088 & 0.121 & 0.084 & 0.106 & 0.026 & 0.010 & 0.056 & 0.081 & 0.007 & 0.009 & 0.059 & 0.072 & 0.034 & 0.005 & 0.035 & 0.095 \\
  clr(Ti) & 0.050 & 0.034 & 0.101 & 0.089 & 0.035 & 0.031 & 0.193 & 0.055 & 0.050 & 0.032 & 0.027 & 0.019 & 0.102 & 0.016 & 0.082 & 0.054 & 0.030 \\
  clr(V) & 0.182 & 0.085 & 0.025 & 0.012 & 0.085 & 0.143 & 0.059 & 0.084 & 0.103 & 0.041 & 0.008 & 0.001 & 0.084 & 0.002 & 0.007 & 0.029 & 0.050 \\
  clr(Y) & 0.002 & 0.042 & 0.083 & 0.189 & 0.019 & 0.085 & 0.102 & 0.063 & 0.023 & 0.068 & 0.025 & 0.035 & 0.063 & 0.036 & 0.059 & 0.073 & 0.034 \\
  clr(Zn) & 0.146 & 0.003 & 0.163 & 0.091 & 0.025 & 0.075 & 0.137 & 0.068 & 0.073 & 0.051 & 0.042 & 0.011 & 0.025 & 0.013 & 0.020 & 0.049 & 0.008 \\
  clr(Zr) & 0.131 & 0.245 & 0.032 & 0.164 & 0.038 & 0.031 & 0.121 & 0.036 & 0.005 & 0.076 & 0.045 & 0.006 & 0.001 & 0.005 & 0.003 & 0.057 & 0.003 \\
   \hline
   Average & 0.114 & 0.099 & 0.093 & 0.089 & 0.060 & 0.060 & 0.058 & 0.054 & 0.047 & 0.046 & 0.044 & 0.043 & 0.042 & 0.041 & 0.037 & 0.037 & 0.035 \\
   \hline
\end{tabular}}}
\end{table}

\begin{table}[ht]
\centering
\red{
\caption{Scaled MASHAP values for SNSS method's mixing function estimate.}\label{table:snss_shap}
\resizebox{\columnwidth}{!}{
\begin{tabular}{rrrrrrrrrrrrrrrrrr}
  \hline
 & IC2 & IC5 & IC7 & IC3 & IC4 & IC6 & IC8 & IC1 & IC14 & IC9 & IC10 & IC16 & IC12 & IC17 & IC11 & IC15 & IC13 \\
  \hline
clr(Al) & 0.096 & 0.216 & 0.017 & 0.003 & 0.015 & 0.063 & 0.007 & 0.165 & 0.040 & 0.068 & 0.079 & 0.062 & 0.031 & 0.066 & 0.011 & 0.011 & 0.051 \\
  clr(Ba) & 0.150 & 0.095 & 0.034 & 0.026 & 0.093 & 0.027 & 0.046 & 0.022 & 0.053 & 0.030 & 0.134 & 0.037 & 0.088 & 0.064 & 0.061 & 0.010 & 0.029 \\
  clr(Ca) & 0.102 & 0.263 & 0.026 & 0.041 & 0.046 & 0.032 & 0.007 & 0.068 & 0.091 & 0.067 & 0.019 & 0.030 & 0.025 & 0.077 & 0.009 & 0.064 & 0.035 \\
  clr(Cr) & 0.131 & 0.037 & 0.068 & 0.019 & 0.186 & 0.128 & 0.043 & 0.089 & 0.059 & 0.035 & 0.064 & 0.002 & 0.027 & 0.037 & 0.004 & 0.041 & 0.030 \\
  clr(Fe) & 0.196 & 0.046 & 0.027 & 0.030 & 0.062 & 0.056 & 0.018 & 0.056 & 0.139 & 0.023 & 0.082 & 0.002 & 0.055 & 0.049 & 0.014 & 0.042 & 0.104 \\
  clr(K) & 0.112 & 0.107 & 0.024 & 0.113 & 0.072 & 0.022 & 0.061 & 0.060 & 0.037 & 0.073 & 0.073 & 0.016 & 0.076 & 0.063 & 0.021 & 0.041 & 0.028 \\
  clr(Mg) & 0.238 & 0.111 & 0.029 & 0.053 & 0.021 & 0.002 & 0.157 & 0.073 & 0.026 & 0.021 & 0.114 & 0.016 & 0.046 & 0.001 & 0.008 & 0.039 & 0.045 \\
  clr(Mn) & 0.070 & 0.034 & 0.040 & 0.047 & 0.041 & 0.071 & 0.056 & 0.099 & 0.165 & 0.029 & 0.013 & 0.054 & 0.110 & 0.063 & 0.030 & 0.062 & 0.016 \\
  clr(Na) & 0.005 & 0.144 & 0.089 & 0.124 & 0.137 & 0.086 & 0.063 & 0.080 & 0.062 & 0.014 & 0.021 & 0.099 & 0.008 & 0.017 & 0.021 & 0.019 & 0.012 \\
  clr(Nb) & 0.078 & 0.082 & 0.100 & 0.099 & 0.011 & 0.089 & 0.099 & 0.001 & 0.017 & 0.080 & 0.004 & 0.070 & 0.067 & 0.012 & 0.116 & 0.053 & 0.021 \\
  clr(P) & 0.113 & 0.057 & 0.083 & 0.029 & 0.028 & 0.058 & 0.004 & 0.081 & 0.085 & 0.145 & 0.000 & 0.169 & 0.023 & 0.039 & 0.049 & 0.028 & 0.008 \\
  clr(Si) & 0.289 & 0.082 & 0.044 & 0.069 & 0.043 & 0.082 & 0.057 & 0.003 & 0.030 & 0.119 & 0.025 & 0.011 & 0.064 & 0.017 & 0.045 & 0.019 & 0.001 \\
  clr(Sr) & 0.003 & 0.152 & 0.044 & 0.112 & 0.093 & 0.031 & 0.040 & 0.064 & 0.032 & 0.080 & 0.084 & 0.045 & 0.081 & 0.058 & 0.071 & 0.004 & 0.008 \\
  clr(Ti) & 0.050 & 0.106 & 0.131 & 0.001 & 0.053 & 0.070 & 0.114 & 0.014 & 0.089 & 0.031 & 0.133 & 0.073 & 0.027 & 0.066 & 0.031 & 0.005 & 0.005 \\
  clr(V) & 0.153 & 0.016 & 0.068 & 0.047 & 0.072 & 0.097 & 0.075 & 0.030 & 0.057 & 0.033 & 0.089 & 0.006 & 0.000 & 0.056 & 0.063 & 0.065 & 0.072 \\
  clr(Y) & 0.079 & 0.071 & 0.151 & 0.191 & 0.018 & 0.000 & 0.106 & 0.030 & 0.006 & 0.042 & 0.018 & 0.011 & 0.055 & 0.049 & 0.048 & 0.094 & 0.028 \\
  clr(Zn) & 0.086 & 0.031 & 0.154 & 0.164 & 0.126 & 0.157 & 0.064 & 0.059 & 0.051 & 0.053 & 0.003 & 0.009 & 0.003 & 0.015 & 0.018 & 0.003 & 0.004 \\
  clr(Zr) & 0.328 & 0.063 & 0.137 & 0.092 & 0.028 & 0.000 & 0.050 & 0.063 & 0.001 & 0.039 & 0.017 & 0.092 & 0.006 & 0.026 & 0.031 & 0.008 & 0.019 \\
   \hline
   Average & 0.127 & 0.095 & 0.070 & 0.070 & 0.064 & 0.060 & 0.059 & 0.059 & 0.058 & 0.055 & 0.054 & 0.045 & 0.044 & 0.043 & 0.036 & 0.034 & 0.029 \\
   \hline
\end{tabular}}}
\end{table}

\begin{figure}
  \begin{minipage}[b]{0.5\linewidth}
    \centering
    \includegraphics[width=.9\linewidth]{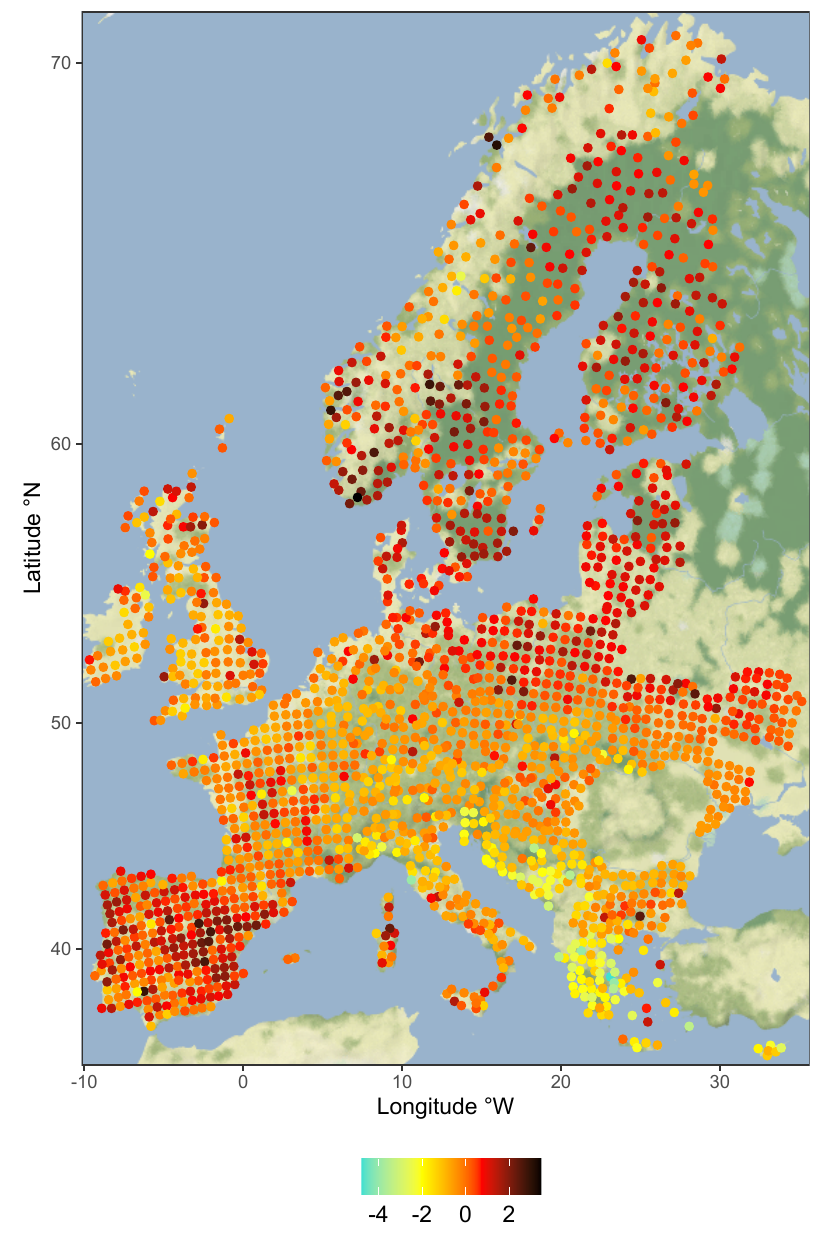} 
  \end{minipage}
  \begin{minipage}[b]{0.5\linewidth}
    \centering
    \includegraphics[width=.9\linewidth]{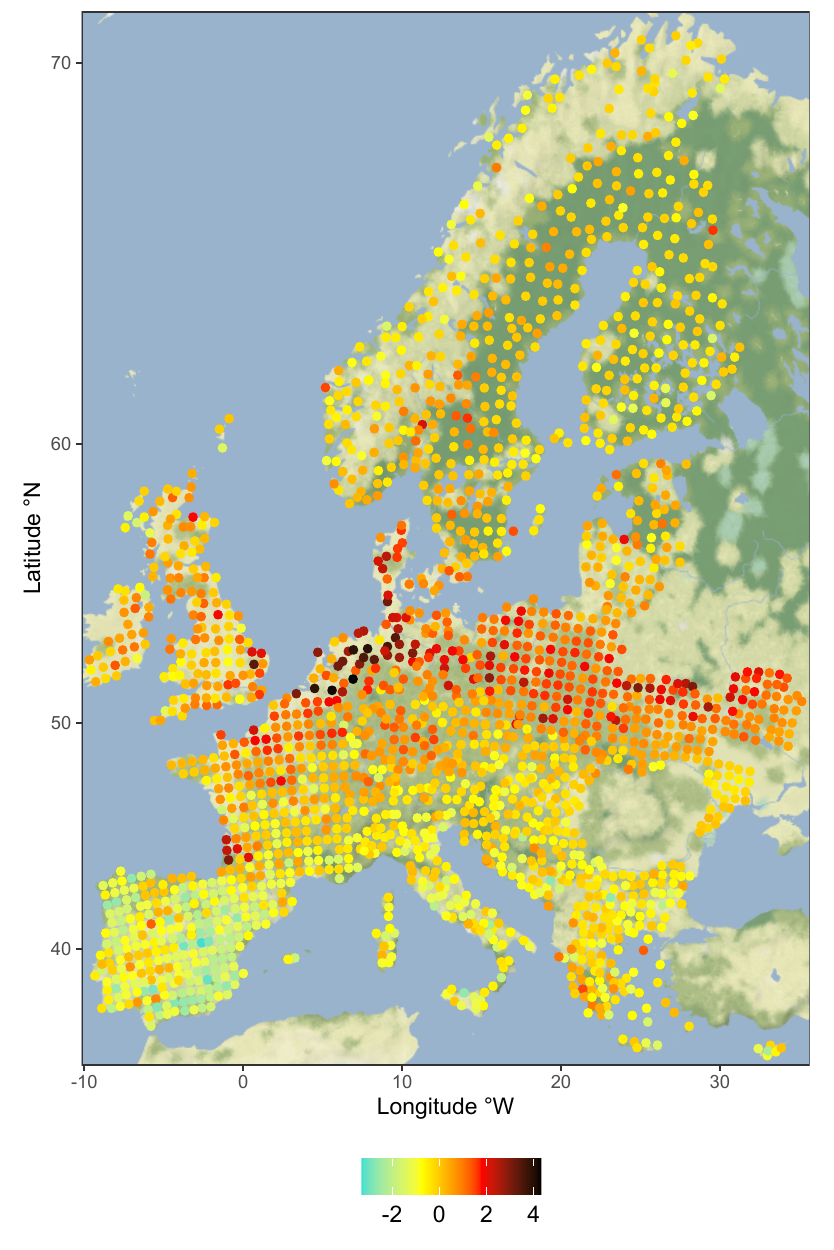} 
  \end{minipage}
\caption{\red{The first (left) and the second (right) latent fields for GEMAS dataset recovered by SBSS.}}\label{fig:ics_sbss}
\end{figure}

\begin{figure}
  \begin{minipage}[b]{0.5\linewidth}
    \centering
    \includegraphics[width=.9\linewidth]{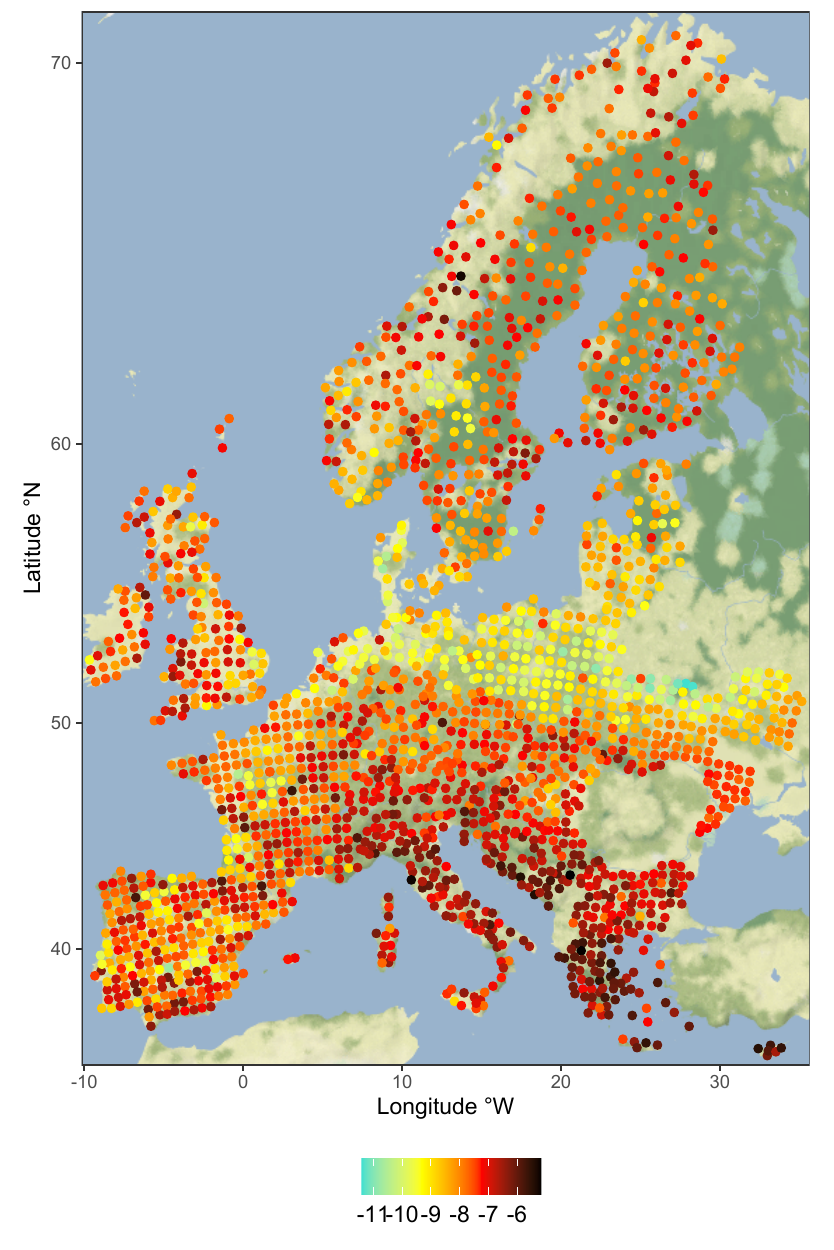} 
  \end{minipage}
  \begin{minipage}[b]{0.5\linewidth}
    \centering
    \includegraphics[width=.9\linewidth]{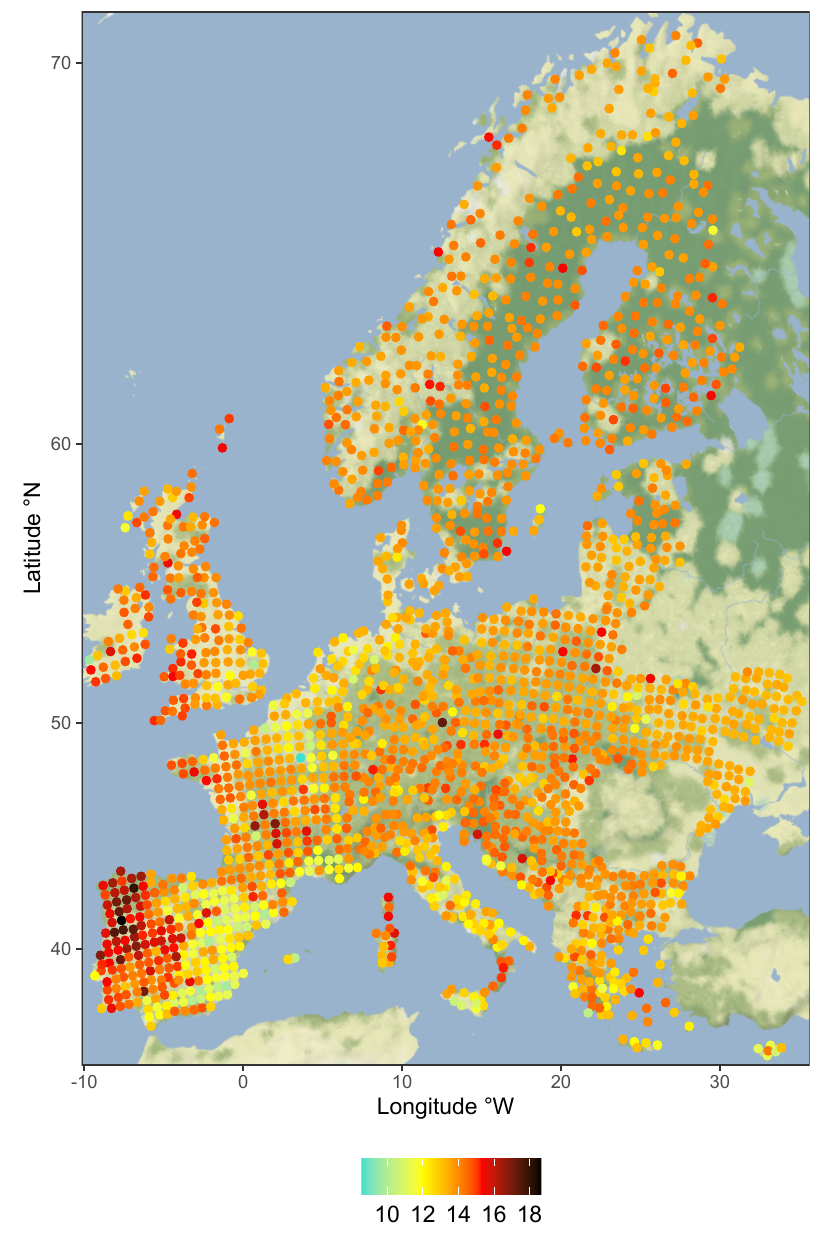} 
  \end{minipage}
\caption{\red{The first (left) and the second (right) latent fields for GEMAS dataset recovered by SNSS.}}\label{fig:ics_snss}
\end{figure}

\begin{table}
\centering
\caption{The scaled MASHAP values for the iVAE model's encoder part.}\label{table:SHAP_encoder}
\resizebox{\columnwidth}{!}{
\begin{tabular}{rrrrrrrrrrrrrrrrrr}
  \hline
 & IC1 & IC2 & IC3 & IC4 & IC5 & IC6 & IC7 & IC8 & IC9 & IC10 & IC11 & IC12 & IC13 & IC14 & IC15 & IC16 & IC17 \\
  \hline
clr(Al) & 0.042 & 0.027 & 0.024 & 0.014 & 0.042 & 0.039 & 0.050 & 0.011 & 0.109 & 0.017 & 0.008 & 0.082 & 0.009 & 0.036 & 0.038 & 0.042 & 0.036 \\
  clr(Ba) & 0.128 & 0.047 & 0.017 & 0.029 & 0.086 & 0.009 & 0.142 & 0.037 & 0.201 & 0.018 & 0.060 & 0.095 & 0.059 & 0.060 & 0.059 & 0.069 & 0.057 \\
  clr(Ca) & 0.057 & 0.221 & 0.028 & 0.089 & 0.039 & 0.155 & 0.114 & 0.042 & 0.063 & 0.090 & 0.133 & 0.028 & 0.038 & 0.066 & 0.075 & 0.070 & 0.085 \\
  clr(Cr) & 0.029 & 0.031 & 0.013 & 0.084 & 0.014 & 0.063 & 0.028 & 0.155 & 0.031 & 0.134 & 0.027 & 0.027 & 0.062 & 0.058 & 0.053 & 0.065 & 0.058 \\
  clr(Fe) & 0.055 & 0.015 & 0.028 & 0.043 & 0.033 & 0.074 & 0.056 & 0.019 & 0.019 & 0.013 & 0.089 & 0.128 & 0.014 & 0.039 & 0.045 & 0.053 & 0.048 \\
  clr(K) & 0.062 & 0.006 & 0.062 & 0.103 & 0.031 & 0.008 & 0.007 & 0.059 & 0.015 & 0.070 & 0.023 & 0.087 & 0.015 & 0.060 & 0.059 & 0.066 & 0.052 \\
  clr(Mg) & 0.072 & 0.110 & 0.161 & 0.068 & 0.129 & 0.086 & 0.015 & 0.035 & 0.019 & 0.046 & 0.029 & 0.025 & 0.171 & 0.073 & 0.097 & 0.064 & 0.068 \\
  clr(Mn) & 0.042 & 0.024 & 0.125 & 0.006 & 0.034 & 0.098 & 0.215 & 0.027 & 0.058 & 0.008 & 0.025 & 0.045 & 0.032 & 0.058 & 0.054 & 0.051 & 0.051 \\
  clr(Na) & 0.057 & 0.086 & 0.043 & 0.057 & 0.156 & 0.069 & 0.038 & 0.017 & 0.038 & 0.079 & 0.101 & 0.022 & 0.076 & 0.068 & 0.053 & 0.062 & 0.057 \\
  clr(Nb) & 0.015 & 0.052 & 0.019 & 0.026 & 0.027 & 0.016 & 0.007 & 0.074 & 0.050 & 0.086 & 0.036 & 0.130 & 0.024 & 0.052 & 0.053 & 0.050 & 0.069 \\
  clr(P) & 0.044 & 0.009 & 0.134 & 0.072 & 0.067 & 0.087 & 0.192 & 0.051 & 0.007 & 0.020 & 0.031 & 0.032 & 0.015 & 0.053 & 0.051 & 0.053 & 0.047 \\
  clr(Si) & 0.078 & 0.033 & 0.007 & 0.016 & 0.057 & 0.063 & 0.011 & 0.141 & 0.101 & 0.042 & 0.071 & 0.011 & 0.086 & 0.066 & 0.050 & 0.034 & 0.064 \\
  clr(Sr) & 0.051 & 0.102 & 0.068 & 0.035 & 0.057 & 0.010 & 0.017 & 0.068 & 0.032 & 0.071 & 0.160 & 0.008 & 0.034 & 0.047 & 0.049 & 0.057 & 0.054 \\
  clr(Ti) & 0.028 & 0.054 & 0.043 & 0.046 & 0.055 & 0.012 & 0.026 & 0.087 & 0.044 & 0.044 & 0.074 & 0.007 & 0.051 & 0.056 & 0.056 & 0.065 & 0.063 \\
  clr(V) & 0.093 & 0.018 & 0.019 & 0.054 & 0.057 & 0.038 & 0.011 & 0.035 & 0.095 & 0.151 & 0.020 & 0.017 & 0.057 & 0.048 & 0.050 & 0.051 & 0.043 \\
  clr(Y) & 0.012 & 0.055 & 0.067 & 0.030 & 0.087 & 0.034 & 0.019 & 0.061 & 0.104 & 0.017 & 0.029 & 0.072 & 0.136 & 0.054 & 0.046 & 0.048 & 0.058 \\
  clr(Zn) & 0.059 & 0.039 & 0.112 & 0.166 & 0.012 & 0.115 & 0.042 & 0.063 & 0.007 & 0.064 & 0.026 & 0.048 & 0.011 & 0.052 & 0.050 & 0.051 & 0.040 \\
  clr(Zr) & 0.077 & 0.070 & 0.032 & 0.064 & 0.018 & 0.026 & 0.009 & 0.017 & 0.007 & 0.030 & 0.060 & 0.134 & 0.110 & 0.053 & 0.063 & 0.049 & 0.050 \\
   \hline
\end{tabular}
}
\end{table}

\end{document}